# Enhanced Geological Prediction for Tunnel Excavation Using Full Waveform Inversion Integrating Sobolev Space Regularization with a Quadratic Penalty Method

**Jiahang Li [a,b,*], Junichi Takekawa [a], Keisuke Kurihara [c], Karnallisa Desmy Halim [c], Kazuhiko Masumoto [c], Yasuyuki Miyajima [c]**

[a]*Department of Civil and Earth Resources Engineering, Graduate School of Engineering, Kyoto University, Kyoto 615-8246, Japan*

[b]*Geoscience Enterprise Inc. (GSE), Tokyo 105-0011, Japan*

[c]*Kajima Technical Research Institute, Kajima Corporation, Tokyo 182-0036, Japan*

[*]***Corresponding Author**: **Jiahang Li***

[a]*Department of Civil and Earth Resources Engineering, Kyoto University, Kyoto 615-8246, Japan.*

[b]*Geoscience Enterprise Inc. (GSE), Tokyo 105-0011, Japan.*

*Email address: lijiahang0921@gmail.com*

## ABSTRACT

In the process of tunnel excavation, advanced geological prediction technology has become indispensable for safe, economical, and efficient tunnel construction. Although traditional methods such as drilling and geological analysis are effective, they typically involve destructive processes, carry high risks, and incur significant costs. In contrast, non-destructive geophysical exploration offers a more convenient and economical alternative. However, the accuracy and precision of these non-destructive methods can be severely compromised by complex geological structures, restrictions on observation coverage and environmental noise. To address these challenges effectively, a novel approach using frequency domain full waveform inversion, based on a penalty method and Sobolev space regularization, has been proposed to enhance the performance of non-destructive predictions. The proposed method constructs a soft-constrained optimization problem by restructuring the misfit function into a combination of data misfit and wave equation drive terms to enhance convexity. Additionally, it semi-extends the search space to both the wavefield and the model parameters to mitigate the strong nonlinearity of the optimization, facilitating high-resolution inversion. Furthermore, a Sobolev space regularization algorithm is introduced to flexibly adjust the regularization path, addressing issues related to noise and artefacts to enhance the robustness of the algorithm. We evaluated the performance of the proposed full waveform inversion using several tunnel models with fault structures by comparing the results of the enhanced method with those of traditional least-squares-based Tikhonov regularization and total variation regularization full waveform inversion methods. The verification results confirm the superior capabilities of the proposed method as expected.

## 1. Introduction

The construction of underground infrastructures has become increasingly comprehensive and influential due to population growth and urbanization. Nearly all underground construction projects involve constructing tunnels, so it's crucial to enhance the safety and reliability of tunnel construction processes (Li et al., 2017). During underground excavation, particularly in tunnel excavation, the presence of potential hazards such as fractured zones, faults, and aquifers ahead of the tunnel poses significant safety risks (Zhang et al., 2018). Therefore, before and during construction, geophysical exploration and measurements of the geological conditions and underground structures at the site are necessary to provide general geological information about the tunnel area. Hence, employing a non-destructive tunnel advance exploration technique to predict unidentified geological structures ahead of the tunnel is of paramount importance (Xu et al., 2021).

Geophysical exploration is a primary method for detecting and identifying underground geological structures. In the past several decades, many scholars have developed exploration methods such as tunnel seismic prediction (TSP), true reflection tomography (TRT), horizontal seismic profiling (HSP), tunnel seismic tomography (TST), tunnel seismic while drilling (TSWD), and tunnel geology prediction (TGP) (Hanson et al., 2002; Petronio et al., 2007; Cheng et al., 2014; Yu et al., 2022). Ashida (2002) proposed to use of three-component seismic records to enhance the imaging capability by incorporating the incident directions of seismic waves. However, the imaging accuracy of these methods is limited because the observation systems are confined to tunnel environments, i.e., observation equipment can only be installed inside the tunnel. This deployment leads to incomplete seismic data collection that fails to meet the requirements for high-precision imaging (Yu et al., 2022). Further, these methods share the same limitations as conventional seismic exploration methods; they do not utilize all seismic wave information, which could lead to the potential neglect of critical seismic information during the process of separating multiple waves.

A noteworthy method in the realm of geophysical exploration is full waveform inversion (FWI), a procedure that surpasses the Born approximation, first proposed by Tarantola (1984). Since FWI can provide high-resolution and highly accurate inversion results, the application of FWI has been promoted mainly in the oil industry. FWI includes intrinsic problems such as cycle-skipping and strong nonlinearity especially when low signal-to-noise ratio, missing data, or insufficient coverage of measurements. To tackle these problems, many novel approaches have been proposed to improve the inversion accuracy even when the observed data is partly missing or affected by noise (Lin et al., 2023; Li et al., 2024a, 2024b). The application of these latest techniques would greatly improve the imaging accuracy ahead of the tunnel face.

In recent years, FWI has started to be applied to estimate geological risks ahead of the tunnel face. Musayev et al. (2016) first applied frequency-domain FWI to tunnel detection in 2016 and discussed its feasibility in imaging velocity fields. Nguyen and Nestorovic (2018) proposed an enhanced tunnel seismic wave FWI method to locate anomalous areas ahead of the tunnel face. Additionally, Bharadwaj et al. (2017) introduced an FWI approach combining the adjoint gradient method with a two-dimensional model, focusing on horizontally polarized shear waves. Furthermore, for tunnel seismic advance prediction, Riedel et al. (2021) and Riedel et al. (2023) introduced time-domain and frequency-domain elastic wave FWI methods. Although some progress has been made in this area, there remains extensive research space and application prospects for utilizing FWI to predict tunnel excavation processes in advance (Trapp and Nestorović, 2022).

In the present study, we propose to apply an advanced regularization technique and an enhanced optimization algorithm to the tunnel advanced prediction circumstance. Specifically, considering the spatial constraints within the tunnel environment, we deploy an external acquisition system placed outside the tunnel, with many receivers positioned on the free surface of the overlying rock mass. Additionally, we suggest using a Sobolev space regularization algorithm to config a regularization term in the misfit function, enhancing the noise resistance and artefact suppression capability of tunnel seismic advance prediction FWI. This paper compares and discusses the multi-scale frequency-domain quadratic penalty Sobolev space regularization FWI algorithm to least-squares (LS) based on other regularization algorithms including Tikhonov and total variation (TV) regularization. And the analyzing of the performance improvement by enhanced adjoint-state variables, wavefield calculations and misfit functions algorithm is also the nucleus of this paper. Lastly, the effectiveness of the proposed

FWI on tunnel seismic advance prediction is verified through the synthetic tunnel models with different types of faults, and noise conditions.

## 2. Methodology

In FWI, a reasonable forward calculation is crucial for inversion, as accurate inversion is based on correct forward modelling (Virieux and Operto, 2009). In geophysical exploration techniques, frequency domain FWI has attracted widespread interest and research for the exceptional accuracy and flexibility it offers, which are not constrained by overly idealized sampling rates in time (Takekawa and Mikada, 2018).

### 2.1. Source-receiver frequency domain modelling

A critical component of the forward method in the frequency domain is solving the Helmholtz equation (Huang et al., 2017), which is an important math structure for comprehending geophysical phenomena. For geophysical problems, the following equation applies to the frequency-domain wave equation, particularly when considering complex media and source configurations:

$$(\Delta + \omega^2\mathbf{m})\mathbf{u} = -L(\omega)\delta(\mathbf{x} - \mathbf{x}_s)\delta(\mathbf{z} - \mathbf{z}_s), \quad \mathbf{(1)}$$

where $\Delta$ denotes the Laplacian operator, $\omega$ denotes the angular frequency, $\mathbf{m}$ represents the squared slowness, $\mathbf{u}$ denotes the wavefield, $L(\omega)$ represents the frequency-dependent linear operator, $\delta(\mathbf{x}-\mathbf{x}_s)$ and $\delta(\mathbf{z}-\mathbf{z}_s)$ denote the source at points $\mathbf{x}_s$ and $\mathbf{z}_s$ (Habashy et al., 2011). Frequency-domain FWI minimizes the LS difference between observed and computed data in the data space, then iteratively updates the model based on gradient direction and model increments and back to the model space, which means the whole process heavily relies on the accuracy of the forward modelling. The frequency-domain method, noted for its high precision in handling low-frequency components and its ability to avoid local minima, compared to the time-domain, frequency-domain algorithms are usually more efficient because they can be solved directly in frequency space, avoiding the accumulation of errors in time stepping (Ovcharenko et al., 2019). These advantages make frequency-domain FWI particularly beneficial for advance prediction in tunnel excavation.

### 2.2. A soft-constrained optimization

The inversion problem of the frequency-domain FWI can be formulated as a nonlinear partial differential equation (PDE)-constrained optimization problem (Brossier et al., 2009):

$$\min_{\mathbf{m},\mathbf{u}} \|\mathbf{Pu} - \mathbf{D}\|_2^2, \quad \text{s.t.} \ \mathbf{A}(\mathbf{m})\mathbf{u} = \mathbf{S}, \tag{2}$$

where the $\mathbf{P}$ denotes observation operator, $\mathbf{D}$ denotes observed data; the right side is subject to a constraint function, where $\mathbf{A}(\mathbf{m})$ represents discretized PDE, $\mathbf{S}$ denotes the source term and $\|\bullet\|$ represents the Euclidean norm (Aghamiry et al., 2019):

$$\mathbf{A}(\mathbf{m})\mathbf{u} = \omega^2 \text{diag}(\mathbf{m}) + \Delta, \tag{3}$$

where $\text{diag}(\bullet)$ denotes the diagonalization operation. For nonlinear constrained optimization problems, specifically Eq. **(2)**, the Lagrange multiplier method can be used for solving by Lagrange relaxation (Gholami et al., 2022). However, commonly solving the above problem requires the use of a full-space approach to jointly update three distinct variables, which are wavefield, model parameter and Lagrange multiplier, meaning that even in single-parameter acoustic inversion, it is necessary to solve a large-scale Karush–Kuhn–Tucker (KKT) conditions (Qi and Jiang, 1997). In this context, the matrix operator is a multivariate Hessian matrix, which implies a substantial computational demand (Aghazade et al., 2022). An alternative simplification involves projecting the full space onto the parameter search space, which entails computing the solution to the wave equation within the misfit to enhance computational efficiency. Based on this, the constrained optimization problem of Eq. **(2)** can thus be transformed into the following unconstrained optimization problem:

$$\min_{\mathbf{m}} F(\mathbf{m}) = \min_{\mathbf{m}} \left\| \mathbf{P}\mathbf{A}^{-1}(\mathbf{m})\mathbf{S} - \mathbf{D} \right\|_2^2, \quad \textbf{(4)}$$

the simplified unconstrained form of the misfit function $F(\mathbf{m})$, as expressed above, offers easier manageability compared to full-space computations. However, the inversion challenge intensifies because the PDE inverse operator $\mathbf{A}^{-1}$ exhibits a highly oscillatory nature, which highlights almost all optimization problems' inherent nonlinearity (Gholami and Aghazade, 2024). Consequently, the inversion process of FWI becomes increasingly reliant on an accurate initial model or the incorporation of low-frequency information to avoid falling into local minima.

To overcome the above problems, a soft-constrained quadratic penalty term method concerning the wavefield and model parameters is proposed (Leeuwen and Herrmann, 2015):

$$\min_{\mathbf{m},\mathbf{u}} F(\mathbf{m}) = \min_{\mathbf{m},\mathbf{u}} \left\| \mathbf{P}\mathbf{u} - \mathbf{D} \right\|_2^2 + \tau \left\| \mathbf{A}(\mathbf{m})\mathbf{u} - \mathbf{S} \right\|_2^2, \quad \textbf{(5)}$$

where $\tau$ is a non-negative penalty parameter includes a scaling factor $\gamma$. Relative to conventional least-square algorithms, this method broadens the search space from not only the model parameter but also includes the wave equation. Moreover, unlike full-space constrained optimization, Eq. **(4)** projects the full-space wavefield, model parameters, and the Laplacian operator onto a reduced space comprising only the wavefield and model parameters, which significantly mitigates the nonlinearity inherent in the inversion problem (Li et al., 2024a). The closed-form expression of the wavefield of Eq. **(5)** by the alternating direction method is as follows:

$$\mathbf{u}^{n+1} = \left(\tau \mathbf{A}^T(\mathbf{m}^n)\mathbf{A}(\mathbf{m}^n) + \mathbf{P}^T\mathbf{P}\right)^{-1}\left(\tau \mathbf{A}^T(\mathbf{m}^n)\mathbf{S} + \mathbf{P}\mathbf{D}\right), \quad \textbf{(6)}$$

Eq. **(6)** involves an alternating direction optimization strategy, where the wavefield variable is held fixed while the model parameters are updated. And $\tau \mathbf{A}^T\mathbf{A} + \mathbf{P}^T\mathbf{P}$ involves a simplified version of the original PDE, which means the actual measurements can be used as constraints to auxiliary in adjusting, validating and improving the model (Leeuwen and Herrmann, 2015). Optimization process using a Gauss-Newton iteration to update the model parameters and calculate the gradient concerning them to determine the direction of the next iteration. At this point, it is crucial that the adjoint state $\mathbf{v}$ is completed synchronously with the forward process and does not require an additional solution to the PDE:

$$\mathbf{v}^{n+1} = \tau(\mathbf{A}(\mathbf{m}^{n})\mathbf{u}^{n+1} - \mathbf{S}), \tag{7}$$

where $\mathbf{v}$ is the adjoint-state variable (van Leeuwen and Herrmann, 2013).

Another important concept is the choice of parameters for the penalty term, how to find the appropriate Lagrangian stability point by adjusting the penalty parameter is crucial, the specific derivation process is in Appendix A. We need to find a part of the penalty parameter based on the real part of the largest eigenvalue matrix:

$$\eta = \Re(\Lambda_{\max}(\mathbf{A}^{-T}\mathbf{P}^{T}\mathbf{P}\mathbf{A}^{-1})), \tag{8}$$

where $\Lambda_{\max}$ represents the largest eigenvalue of $\mathbf{A}^{-T}\mathbf{P}^{T}\mathbf{P}\mathbf{A}^{-1}$, $\eta$ is the part of the penalty parameter $\tau$ in Eq. **(5)** (Leeuwen and Herrmann, 2015), which will be analysed and explained in detail in the following part.

## 2.3. Solving procedure with modified Sobolev space regularization-based quadratic penalty method

The gradient of the objective function in FWI typically comprises two main components during the optimization: a data error item and a regularization item. The data error item minimizes the differences between simulated and observed data information, while the regularization term serves to control the smoothness and stability of the solution, prevent overfitting, and address noise and uncertainties in the data information. The influence of Tikhonov regularization (Asnaashari et al., 2013) and TV regularization in inversion problems is well-known (Brandsberg-Dahl et al., 2017); Tikhonov regularization tends to favour smoother models which enhance continuity, whereas TV regularization is known to suffer from the staircase effect (Wang and Feng, 2021). However, more advanced variable regularization methods with adjustable norm paths are rarely considered for adoption in FWI, due to the complexity of the theory. But for practical production in tunnel advance prediction, adjustable regularization that can be adapted to different geological conditions is a better choice. In this paper, based on the reconstruction of wavefields, we consider using a $W_p^1$ norm in Sobolev spaces to balance data fitting accuracy with the physical interpretability of the model, simultaneously enhancing the stability and robustness of the inversion process (Adams and Fournier, 2003; Fischer and Steinwart, 2020). With the addition of the regularization term, Eq. **(5)** can be implicitly expanded as:

$$F(\mathbf{m}) = \Phi(\mathbf{m}) + \Psi(\mathbf{m}), \tag{9}$$

where $\Phi(\mathbf{m})$ is the quadratic penalty objective term with a data error item and a wave equation item based on the quadratic penalty method, $\Psi(\mathbf{m})$ is the regularization term. In the left side term of Eq. **(9)**, the closed-form solution of the eliminated wavefield $\mathbf{u}^{n+1}$ of the quadratic penalty has been shown in Eq. **(6)**. Then, the estimated wavefield is brought back into the optimization, and a new inter-loop of misfit computation procedure can obtain the model parameter variables according to the alternating principle:

$$\begin{aligned}\Phi(\mathbf{m}^n) = &\sum_{i=1}^{N_a}\sum_{j=1}^{N_b}(\mathbf{P}\mathbf{u}_{i,j}^{n+1} - \mathbf{D}_{i,j})^H(\mathbf{P}\mathbf{u}_{i,j}^{n+1} - \mathbf{D}_{i,j}) + \\ &\tau\sum_{i=1}^{N_a}\sum_{j=1}^{N_b}(\mathbf{A}_{i,j}(\mathbf{m}^n)\mathbf{u}_{i,j}^{n+1} - \mathbf{S}_{i,j})^H \times (\mathbf{A}_{i,j}(\mathbf{m}^n)\mathbf{u}_{i,j}^{n+1} - \mathbf{S}_{i,j}),\end{aligned} \tag{10}$$

where $N_a$ denotes the number of discretized frequencies, $N_b$ is the number of sources in a deployment. $(\bullet)^H$ denotes the Hermitian transpose operation (Feng et al., 2022). Eq. **(6)** and Eq. **(10)** ensure that the wavefield computation is not solely based on solving PDE, i.e., it does not rely exclusively on the mathematical methods of solving full PDE (Grote et al., 2014). Instead, it incorporates more extensive analysis and utilization of actual observational data and involves implementing data information drive models' updates to enhance the accuracy of the wavefield estimation. The advantage is that accuracy is guaranteed, and convergence is faster in the case of approximate solutions computed in a computationally acceptable solution space.

Eq. **(9)** on the right side includes a regularization term that is used to control the smoothness and stability of the solution, prevent overfitting, and address noise and uncertainties in the data. Consequently, the original optimization problem can be reconstructed as follows:

$$J(m) = \Phi(\mathbf{m}) + \beta\|\mathbf{m}\|_{w_p^1}, \qquad \textbf{(11)}$$

where $\beta$ is a regularization parametric operator, $\|\bullet\|_{w_p^1}$ is $w_p^1$ norm in Sobolev space. In the $w_p^1$ norm, $p$ is a selectable value that is not less than zero, depending on the physical significance and the required strength of regularization. When $p=1$, the $w_p^1$ norm exhibits characteristics of the TV norm, while $p=2$, it exhibits characteristics of the Tikhonov regularization. If $p$ value lies between these values, particularly when $p=0.5$, it displays features similar to those of the minimum gradient support (MGS) stabilizer (Kazei et al., 2017). We give a multi-view schematic of the visualisation and simulation of the $w_p^1$ norm in **Fig. 1**.

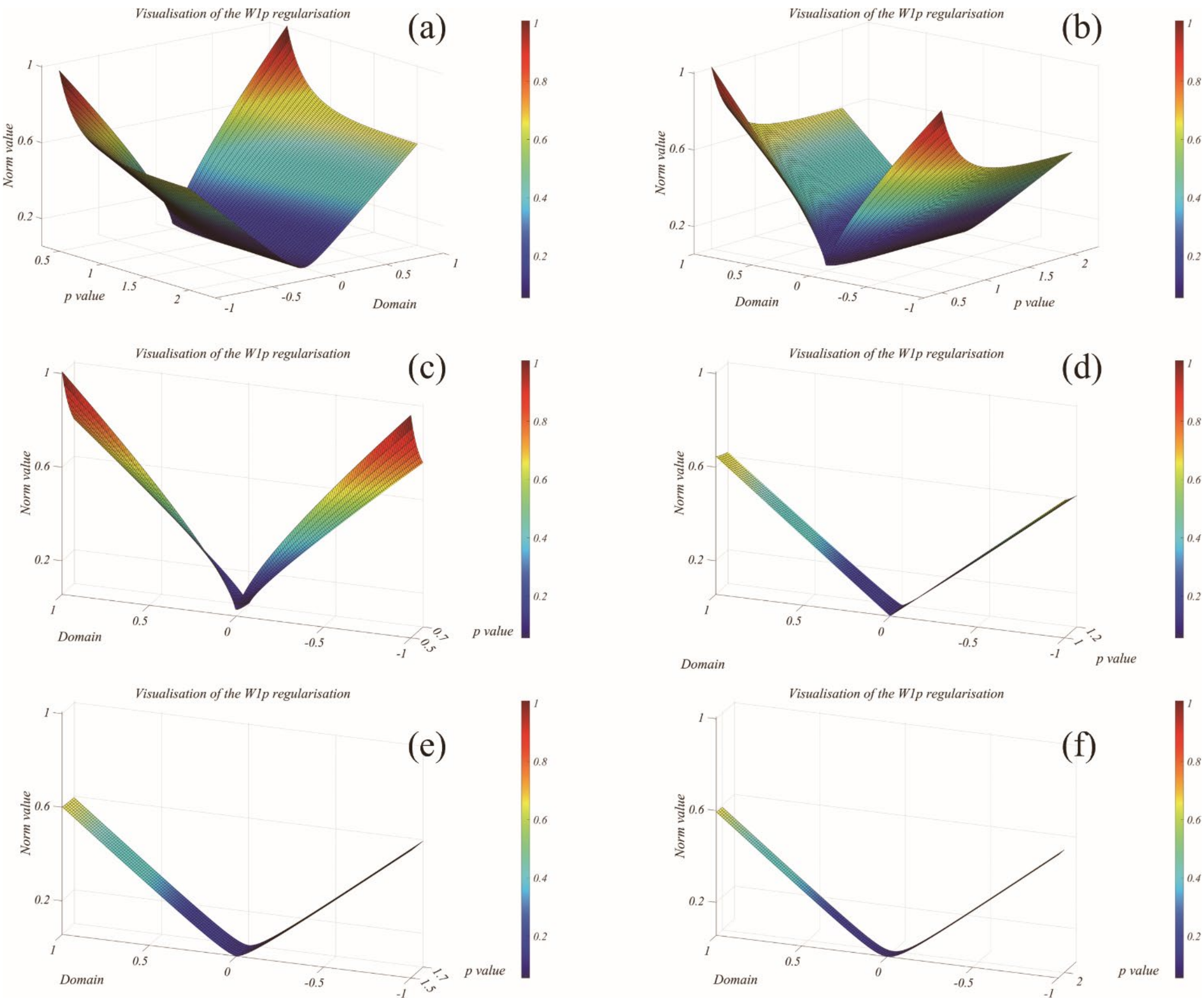

**Fig. 1.** Visualization and simulation of the $W_p^1$ regularization. (a) *p* values ranging from 0.5 to 2, the azimuth angle (AZ) is 50, and the elevation angle (EL) is 15, (AZ: 50°, EL: 15°). (b) *p* values ranging from 0.5 to 2, the AZ is negative 50°, and the elevation angle (EL) is 15°, (AZ: -50°, EL: 15°). (c) *p* values ranging from 0.5 to 0.7, which exhibits the characteristics of the 0.5 norm. (d) *p* values ranging from 1 to 1.2, which exhibits the characteristics of TV regularization. (e) *p* values ranging from 1.5 to 1.7, which exhibits the characteristics of the 1.5 norm. (f) *p* value is equal to 2, which exhibits the characteristics of the Tikhonov regularization.

As an adjustable and flexible norm, the $w_p^1$ norm exhibits various characteristics when $p$

taking different values. This figure illustrates how the $w_p^1$ norm of a two-dimensional vector changes across different $p$ values (ranging from 0.5 to 2) and the domain (varying from -1 to 1). Each $p$ value corresponds to a specific method of norm calculation, and this three-dimensional view demonstrates the dynamic variations of the norm under different parameters. Specifically, **Fig. 1**(a) shows a side view at an azimuth angle of 50 and an elevation angle of 15, whereas **Fig. 1**(b) presents a side view at an azimuth angle of negative 50 and an elevation angle of 15. **Fig. 1**(c)-(f) display side views for $p$ values of 0.5, 1, 1.5, and 2, respectively. It can be observed that as $p$ approaches 2, the $w_p^1$ norm tends towards a convex regularization known as Tikhonov regularization; when $p$ is close to 0.5, the $w_p^1$ norm exhibits a distinct concave. At $p$ equal to 1, the $w_p^1$ norm aligns with TV regularization, remaining convex because a convex function's value on any line segment between two points is less than or equal to the average of the function values at the endpoints. Thus, the $p$ parameter of the $w_p^1$ function maintains its convex property as long as it is greater than or equal to 1, ensuring that the optimization process in FWI stays relatively convex, which helps avoid the impact of local minima.

The revised and optimized formula of the $w_p^1$ norm is as follows:

$$\|\mathbf{m}\|_{w_p^1} = \left(\int_\Gamma |\mathbf{m}(x)|^p \, dx + \int_\Gamma (\nabla\mathbf{m}(x)\cdot\nabla\mathbf{m}(x) \;+\; \sigma)^{p/2} dx\right)^{1/p}, \qquad \textbf{(12)}$$

where $\Gamma$ denotes the model space definition domain, $\mathbf{m}(x)$ denotes the model parameter at position $x$ and $\nabla\mathbf{m}(x)$ denotes the gradient of the model parameter. In Eq. **(12)**, $\sigma$ represents a tiny and positive value for ensuring the numerical stability of the mathematical expressions. Physically, the purpose of adding $\sigma$, which help ensure that calculations remain stable even in regions where changes in the model parameter values are minimal. Thus preventing potential numerical errors or instabilities when the gradient $\nabla\mathbf{m}(x)$ is close to zero. Consequently, the gradient of the Eq. **(11)** can be implicitly expressed as:

$$\nabla_{\mathbf{m}} J(\mathbf{m}) = \nabla_{\mathbf{m}}\Phi(\mathbf{m}) + \beta\nabla_{\mathbf{m}} \|\mathbf{m}\|_{w_p^1}, \qquad \textbf{(13)}$$

Eq. **(13)** integrates the soft constraint containing data error item and wave equation item, and regularization terms, controlling the discrepancy between simulated and observed data as well as the smoothness and complexity of the model. The model parameters are updated via a multi-scale strategy, aiming to prevent overfitting and enhance the physical interpretability of the model. According the Eq. **(3)**, Eq. **(5)** and Eq. **(6)**, the gradient of the misfit penalty term can be expressed as:

$$\nabla_{\mathbf{m}}\Phi(\mathbf{m}) = \sum_{i=1}^{N_a}\sum_{j=1}^{N_b}\omega_{i,j}^2 \mathrm{diag}(\mathbf{u}_{i,j}^{n+1})^T \left(\mathbf{A}_{i,j}(\mathbf{m}^n)\mathbf{u}_{i,j}^{n+1} - \mathbf{S}_{i,j}\right), \tag{14}$$

which computes the gradient of the quadratic penalty across multiple source and receiver configurations. The above equation captures the model's response to various seismic sources and receivers and their discrepancies with observational data. Further, the regularization term's local gradient of the function about the $W_p^1$ norm can also be evaluated as follows:

$$\nabla_{\mathbf{m}}\|\mathbf{m}\|_{W_p^1} = \|\mathbf{m}\|_{W_p^1}^{1-p}\left(p|\mathbf{m}(x)|^{p-2}\mathbf{m}(x) + p\left(\nabla\mathbf{m}(x)\cdot\nabla\mathbf{m}(x)+\sigma\right)^{(p/2)-1}\nabla\mathbf{m}(x)\right), \tag{15}$$

and within this framework, the gradient of the $W_p^1$ norm, and the regularization operator $p$ can be adjusted, which facilitates flexible control over the model's smoothness and numerical stability. Consequently, the proposed method proves particularly effective for identifying geological models that require high continuity and stability, such as faults, especially in scenarios involving the prediction of faults that may be encountered in advance during tunnel excavation processes. In the proposed improved FWI framework, the updating process of the model parameters can be carried out using the alternating minimisation algorithm:

$$\mathbf{m}^{n+1} = \underset{\mathbf{m}_{\min}\le\mathbf{m}\le\mathbf{m}_{\max}}{\arg\min}\ \beta\|\mathbf{m}\|_{W_p^1} + \tau\left\|\omega^2\mathrm{diag}(\mathbf{u}^{n+1}) - \mathbf{S} + \Delta\mathbf{u}^{n+1}\right\|_2^2, \tag{16}$$

this update process allows for the computation and accumulation of wavefields independently and alternating without storing all wavefield information. The system is strictly diagonal, simplifying the computational process.

In summary, the proposed algorithm integrates various benefits of conventional methods and the Lagrangian approach to construct and solve optimization problems, thereby extending the gradient search space. Additionally, the algorithm incorporates a flexible, adjustable regularization method to maintain a balance between sharp and smooth updates in FWI, mitigating nonlinearity and the potential for local minima while enhancing denoising capabilities. We will further elucidate the effectiveness and superiority of the proposed methods through comparison examples in subsequent sections.

## 3. Numerical experiments

The application of FWI on tunnel advance prediction is often limited, dut to the data quality being damaged by the restriction of the deployment of seismic sensors and noise interference in the tunnel environment. In this section, the proposed FWI technique is applied to numerical experiments designed for some special tunnel seismic advance prediction environments to overcome these tough but crucial questions. We build some numerical models (one fault model and two dip-slip faults model) that are difficult to detect using conventional travel-time tomography. We also consider heavy noise interference to reproduce low-quality data of tunnel sites.

### 3.1. Single fault model

#### *3.1.1. Test I: Numerical simulation for single fault model with free surface and with noise*

This paper initially considers a two-dimensional fault model for tunnel advance prediction.

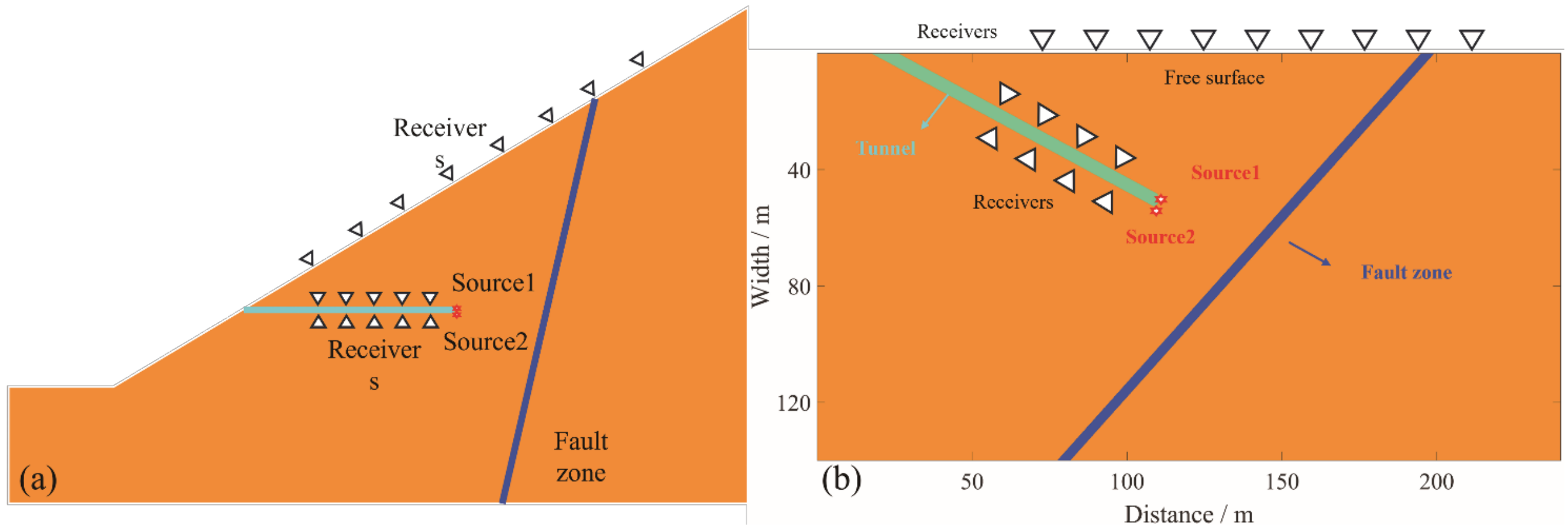

**Fig. 2.** Schematic of the model structure. (a) Tunnel structural model with an inclined fault. (b) The modelling results after tilting the schematic **Fig. 2**(a), the top of the modelling is a free surface to simulate the real ground surface (the number of receivers in the schematic does not represent the real number of receivers).

In **Fig. 2**(a), the model is rotated such that the angle between the tunnel and the free surface is 30 degrees because it is difficult to incorporate an inclined free surface in forward modelling (Takekawa and Mikada, 2019). The tunnel diameter is set at 5 m, the starting position of the tunnel in the model is 20 m from the left side origin, and the velocity is 1.25 km / s. Additionally, a fault structure is located on the right side of the model, with a thickness of 5 m and a fault velocity of 2.5 km / s. The fault starts at a horizontal distance of 80 m. **Fig. 2**(b) illustrates the schematic of the numerical model. The overall size of the model is 240 m × 140 m. The surrounding rock velocity is 3 km / s. In our system configuration, we deploy only two seismic sources positioned in front of the tunnel face like **Fig. 2**(b), without time delay. The number of sources is designed as small as possible to minimize the destructive impact caused by the sources, increase the inversion challenge and simulate extreme conditions for result comparison. Additionally, considering recording multiple reflection waves coming from the free surface, we also place 7 receivers at the tunnel roof together with other 7 receivers at the tunnel bottom. Due to no spatial and environmental restrictions at the free surface of the overlying rock mass, we can bury many receivers; specifically, we place 25 equally spaced receivers, each separated by 7 m.

In this study, we implement absorbing boundary conditions on the left, right, and bottom sides of the two-dimensional isotropic model, while the top surface is treated as a free surface to simulate the actual properties of the geological surface (Takekawa and Mikada, 2016). We utilize acoustic FWI to accurately invert fault structures.

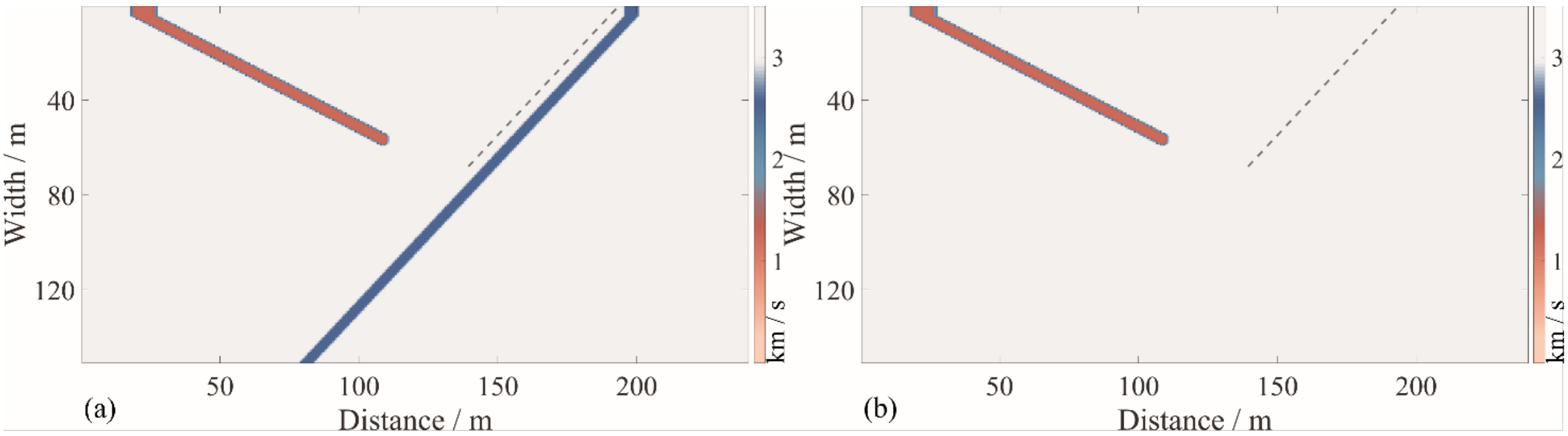

**Fig. 3.** Modelling of a fault structure for tunnel seismic advance prediction. (a) Benchmark of the model with a tilted fault. (b) The initial model of the benchmark. The tunnel structure was removed for inversion self-consistency testing. The grey dotted line is the indicator line.

**Fig. 3**(a) shows the true velocity structure, and **Fig. 3**(b) shows the initial velocity model. In the forward modelling phase, we discretize the model using the finite difference method with a grid length of 6.67 m, comprising a total of 77400 grids. We use a frequency-domain forward modelling algorithm and a multi-scale inversion strategy to mitigate local minima and alleviate nonlinearity. Our frequency stepping approach increases each subsequent inversion frequency by a factor of 1.2, enhancing the efficiency and stability of the inversion process.

Subsequently, we perform a forward process of the benchmark to obtain the real part of the forward wavefield as shown in **Fig. 4**(a). The **Fig. 4**(b) is the imaginary part of the forward wavefield. A forward process of the initial model to obtain the real part of the background wavefield as shown in **Fig. 4**(c). **Fig. 4**(d) is the imaginary part of the initial background wavefield.

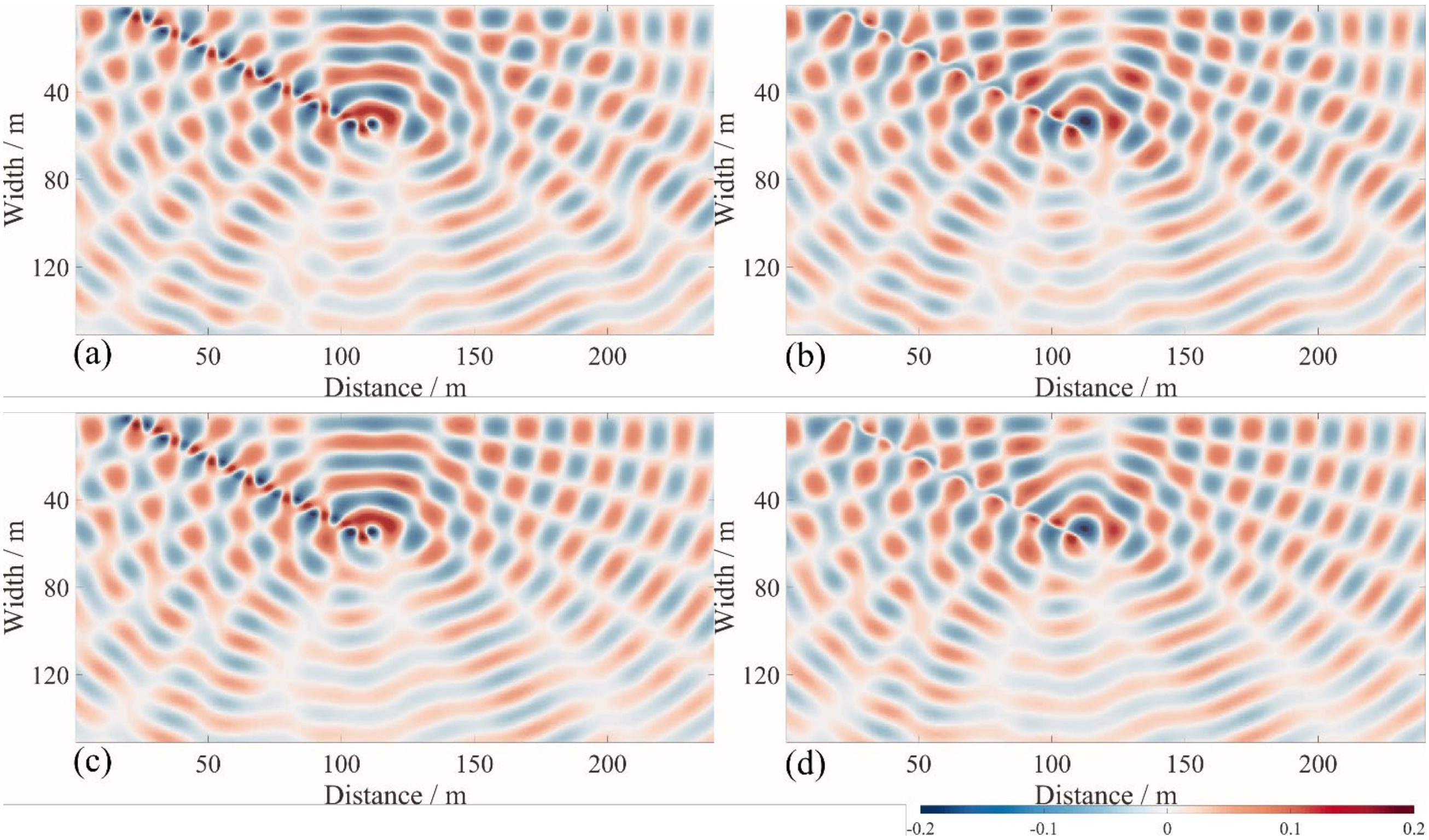

**Fig. 4.** Forward wavefield information. (a) The real part of the forward wavefields. (b) The imaginary part of the forward wavefields. (c) The real part of the initial background wavefield. (d) The imaginary part of the background wavefield.

Another key factor to consider is noise interference in tunnel environments. Particularly in applications of tunnel seismic advance prediction or near-surface surveying, especially in urban or industrial areas, seismic exploration is often affected by vibrations from vehicles, machinery, and other construction activities (Li et al., 2024b). Thus, it is crucial to consider the significant impact of noise on the inversion process. The strength of the artificial noise added to synthetic data is expressed by the following signal-to-noise ratio (SNR):

$$\mathrm{SNR} \ = \ 20 * \mathrm{Log}_{10}\left(\frac{\|\mathbf{D}\|_2}{\|\mathbf{E}\|_2}\right), \qquad \textbf{(17)}$$

where the $\mathbf{E}$ denotes the noise. We consider the impact of adding 4.9 *dB* of random noise to the benchmark data, as shown in **Fig. 5**. **Fig. 5**(a) displays the real part of the clean source-receiver data information using low-frequency source optimized for frequencies of 2 Hz, where the vertical axis indicates the two sources, and the horizontal axis corresponds to 39 receivers.

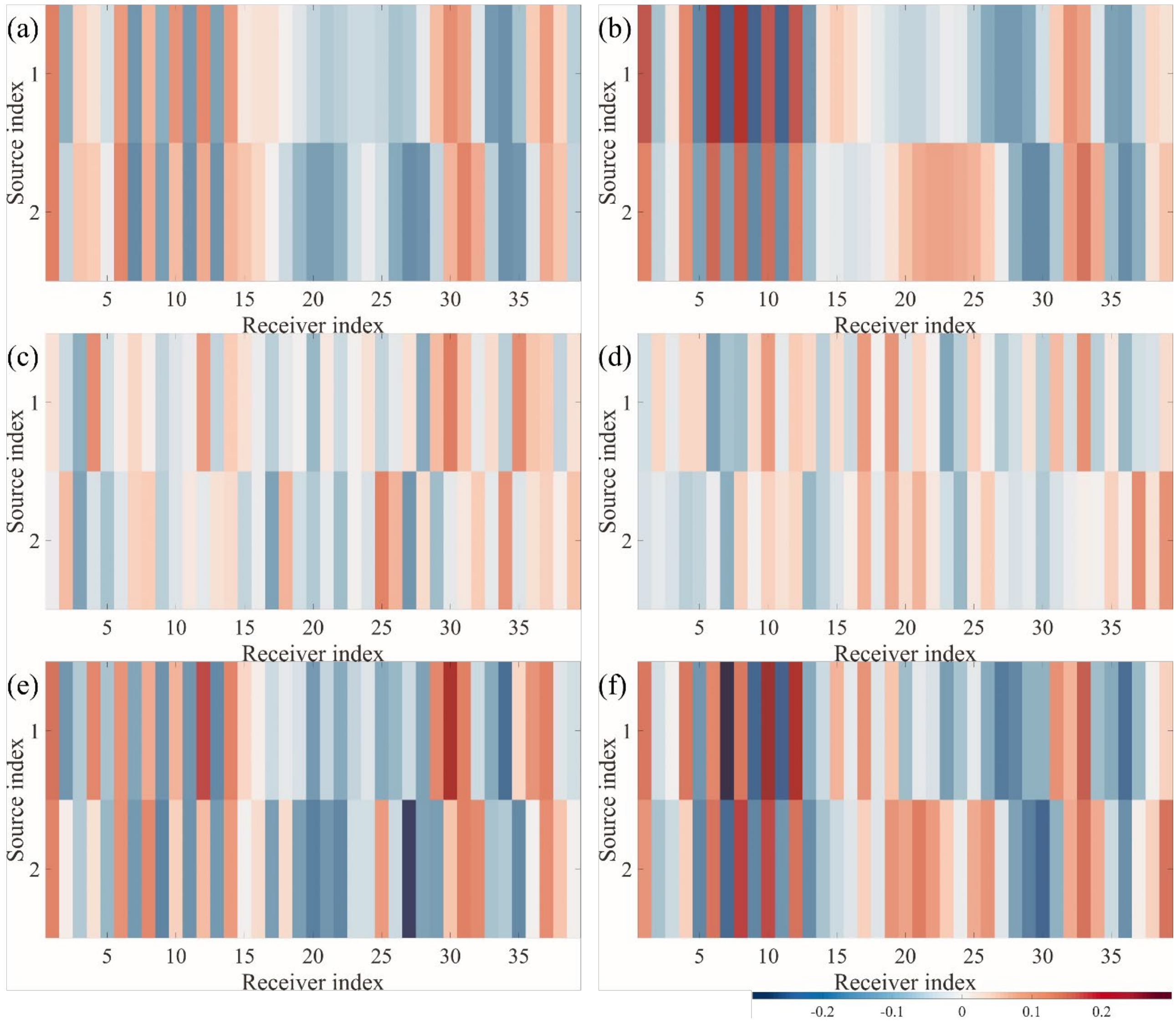

**Fig. 5.** Source-receiver data for the tunnel model. (a) The real part of the clean data without noise component. The vertical coordinate explicit acquisition system consists of two seismic sources, horizontal coordinates from 15 to 39 for receivers on the free surface, and from 1 to 14 for receivers on the tunnel floor and roof. (b) The imaginary part of the clean data without noise component. (c) The real part of the random noise. (d) The imaginary part of the random noise. (e) The real part of the data with random noise as a means for further steps. (f) The imaginary part of the data with random noise as a means for further steps.

Receivers 15-39 are located on the free surface, while receivers 1-14 are positioned on the tunnel floor and roof. **Fig. 5**(b) displays the imaginary part of the clean data. **Fig. 5**(c) illustrates the 4.9 *dB* random noise. **Fig. 5**(d) illustrates the imaginary part of the noise. **Fig. 5**(e) shows the real part of the data information after the noise was added. **Fig. 5**(f) shows the imaginary part of the source-receiver data information after the noise was added.

### *3.1.2. Test I: Optimization*

Before starting the final inversion tests, a critical and non-negligible issue is the selection of the penalty parameter $\tau$ in Eq. **(5)**. As previously mentioned, in the proposed method, the penalty parameter $\tau$ should be represented by the maximum eigenvalue of the matrix $\mathbf{A}^{-T}\mathbf{P}^{T}\mathbf{P}\mathbf{A}^{-1}$ in Eq. **(8)**. However, depending on practical application circumstances, it is necessary to scale the maximum eigenvalue to a certain extent to ensure reaching a suitable magnitude's stationary point of the Lagrangian function within a predefined scaling range (Abdoulaev et al., 2005), because the size of the scaling parameter $\gamma$ directly affects the convergence and stability of the algorithm. Mathematically, as $\gamma$ approaches infinity, the solution of the penalty method tends to satisfy all constraints, but this may also lead to the deterioration of the condition number of the Hessian matrix, making computations unstable (Banerjee et al., 2013). To ensure that the algorithm reaches the expected convergence accuracy within a finite number of iterations, the $\gamma$ value should be small and reason to enforce the constraints while avoiding algorithmic instability. In this section, we will compare and test several different $\gamma$ values to discuss in detail the impact of the penalty parameter on the quadratic penalty method.

Firstly, the performance of four different scaling factors, $\gamma$, is tested under the specific model conditions presented in this paper. **Fig. 6** displays the inversion results for each scaling factor. **Fig. 6**(a) corresponds to $\gamma = 1\mathrm{e}-2$, **Fig. 6**(b) to $\gamma = 1\mathrm{e}-1$, **Fig. 6**(c) to $\gamma = 1\mathrm{e}0$, and **Fig. 6**(d) to $\gamma = 1\mathrm{e}2$.

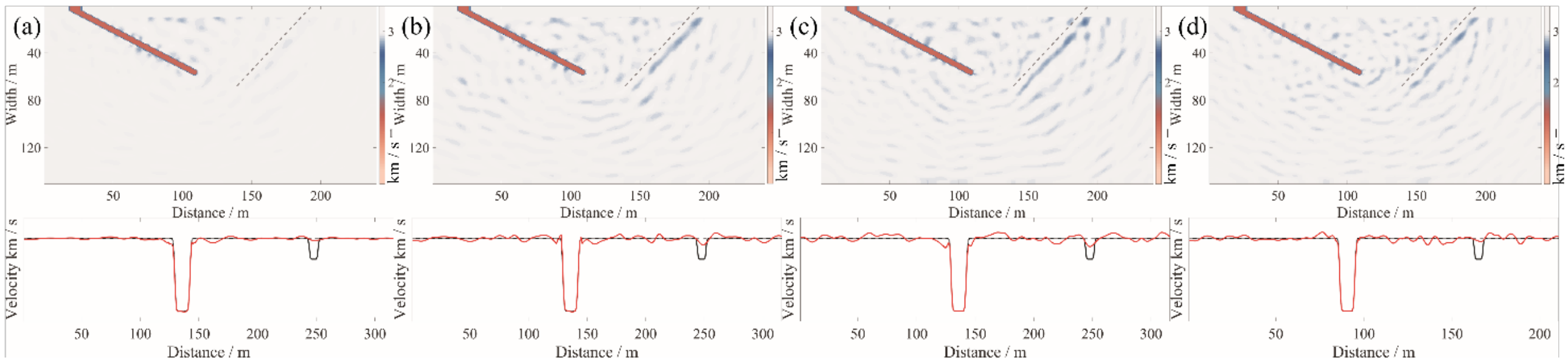

**Fig. 6.** Comparison of inversion tests with different scaling factors $\gamma$ for maximum eigenvalues $\mathbf{A}^{-T}\mathbf{P}^{T}\mathbf{P}\mathbf{A}^{-1}$. (a) $\gamma = 1\mathrm{e}-2$. (b) $\gamma = 1\mathrm{e}-1$. (c) $\gamma = 1\mathrm{e}0$. (d) $\gamma = 1\mathrm{e}2$. Directly below the inversion results, a set of vertical velocity contrasts is displayed, where the black solid line corresponds to the benchmark velocity, the red solid line corresponds to the inverted velocity, and the grey dashed line corresponds to the initial velocity. As the scaling factor increases, leading to a smaller reciprocal of the scaling factor, the inversion results tend to become unstable due to the dominance of the data mismatch term. Conversely, with smaller scaling factors, resulting in a larger reciprocal of the scaling factor, the inversion results are more stable as the wave equation term becomes more predominant. However, if the scaling factor is too small, the data mismatch term will be invalidated.

Theoretically, as the scaling factor increases, the inversion results tend to become unstable due to a smaller reciprocal of the scaling factor, with the data mismatch term becoming dominant. This effect is observed as an increase in artefacts with larger $\gamma$ values. Conversely, smaller scaling factors result in more stable inversion outcomes, where the larger reciprocal enhances the dominance of the wave equation term, however, if the scaling factor is too small, the data mismatch term will be invalidated. The observation from the figures confirms that excessively large scaling factors, or very small reciprocals of $\gamma$, can diminish the influence of the wave equation term, causing the optimization to degrade toward conventional LS. However, excessively small $\gamma$ values may also lead to underfitting.

Thus, subsequent tests will compare both $\gamma = 1\text{e}-1$ and $\gamma = 1\text{e}0$ conditions with conventional algorithms to evaluate their performance. Additionally, the convergence process of the largest eigenvalue of the matrix $\mathbf{A}^{-T}\mathbf{P}^{T}\mathbf{P}\mathbf{A}^{-1}$ under four scaling factor scenarios was compared, as shown in **Fig. 7**.

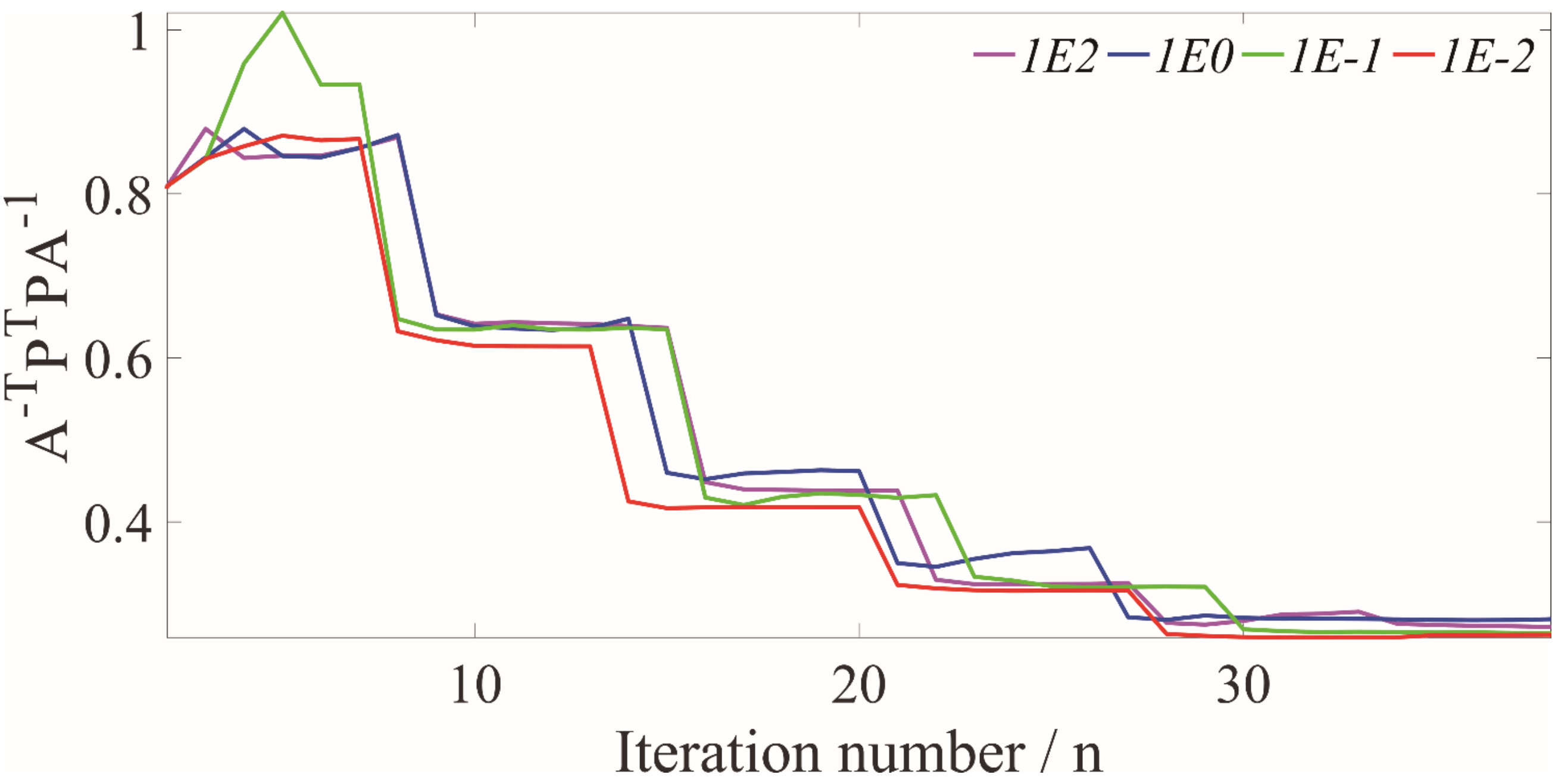

**Fig. 7.** Maximum eigenvalue curve of $\mathbf{A}^{-T}\mathbf{P}^{T}\mathbf{P}\mathbf{A}^{-1}$ in relation to the penalty parameter $\tau$ in Eq. **(8)**. The curves are shown for three different scaling levels, $\gamma = 1\text{e}-2$ in red, $\gamma = 1\text{e}-1$ in green, $\gamma = 1\text{e}0$ in blue and $\gamma = 1\text{e}2$ in magenta.

The adoption of a multi-scale inversion strategy means that the size of the penalty term parameter can significantly influence the smoothness of the convergence process. As observed, smaller scaling factors lead to faster convergence of the $\mathbf{A}^{-T}\mathbf{P}^{T}\mathbf{P}\mathbf{A}^{-1}$ matrix. Furthermore, as derived in Appendix A, faster convergence indicates a higher model update rate.

Additionally, the relative impacts of the data mismatch term and the wave equation term were compared across four scaling factors after 3 iterations, as illustrated in **Fig. 8**.

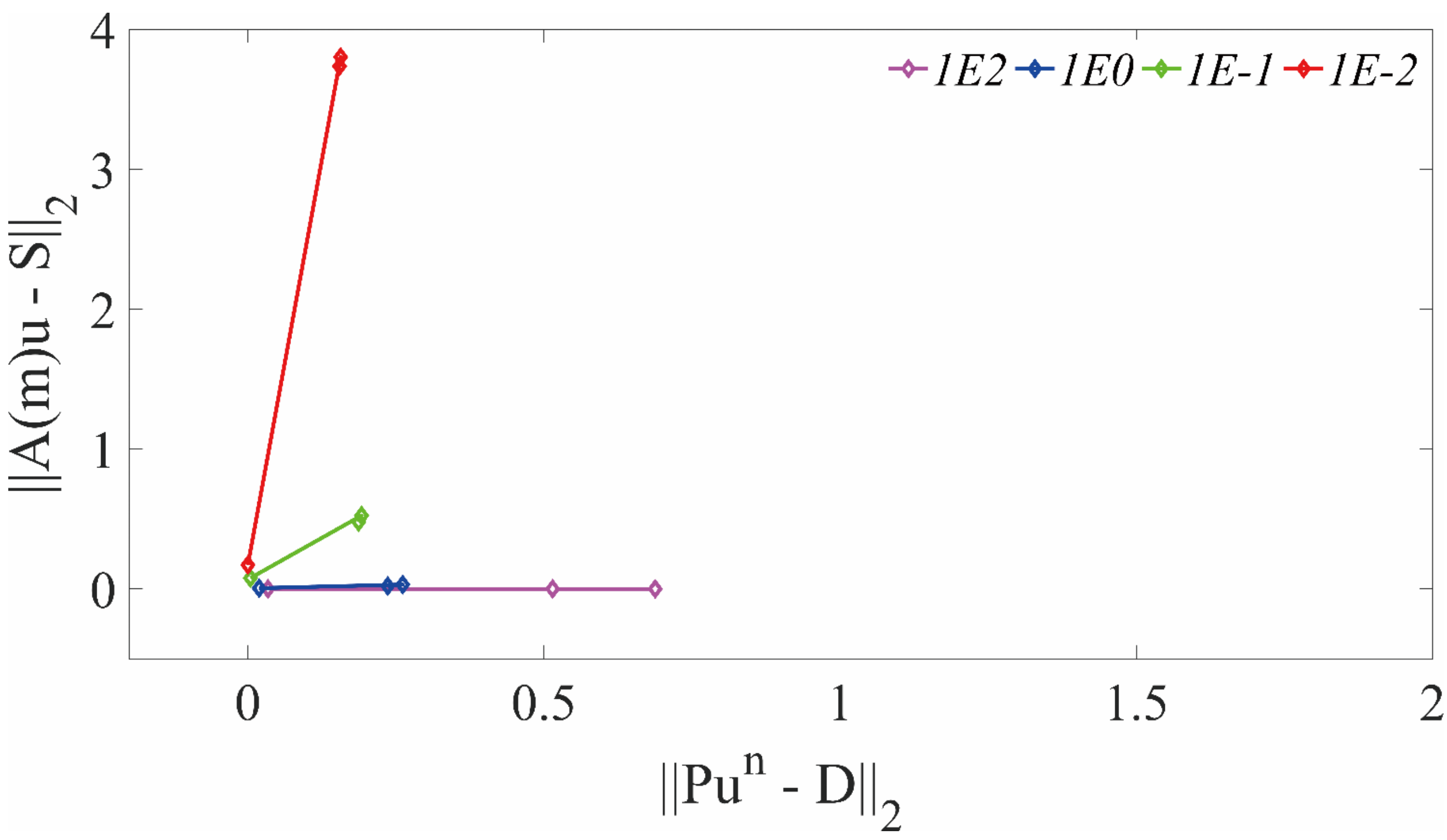

**Fig. 8.** Relative importance of the data mismatch term and the wave equation penalty term in the initial stage of the inversion. The curves are shown for four different scaling levels, $\gamma = 1e-2$ in red, $\gamma = 1e-1$ in green, $\gamma = 1e0$ in blue and $\gamma = 1e2$ in magenta. Smaller scaling factors can ensure that both terms maintain convergent behaviour simultaneously. At $\gamma = 1e-2$, the wavefield term's correlation with the data mismatch term is small, potentially leading to a scenario in which the wave equation item converges independently, suggesting the possibility of the wave equation term falling into a local minimum. At $\gamma = 1e-1$, the wave equation term converges slightly faster than the data term, but these two items are balanced. At $\gamma = 1e0$, the data mismatch term converges faster than the wave equation term, but the data misfit item is dominant, wave equation item is auxiliary. However, at $\gamma = 1e2$, the wavefield term's correlation with the data mismatch term is also too small, potentially leading to a scenario in which the data term converges independently. Importantly, at this scaling, the data mismatch term does not reduce to a level nearly like that of the wave equation term and declines noticeably slower, indicating that the data mismatch term has fallen into a local minimum.

The most significant finding is that reasonable scaling factors ensure that both terms maintain synchronous convergence behaviour. At $\gamma = 1\mathrm{e}-2$, the data mismatch term declines rapidly, but the wave equation term does not reduce to the same level as the data mismatch term, suggesting the possibility of the wave equation term falling into a local minimum. At $\gamma = 1\mathrm{e}-1$, the wave equation term converges slightly faster than the data term, and both decline synchronously. At $\gamma = 1\mathrm{e}0$, the data mismatch term converges faster than the wave equation term, and the overall convergence is faster as $\gamma = 1\mathrm{e}-1$. However, at $\gamma = 1\mathrm{e}2$, the wavefield term's correlation with the data mismatch term is too small, potentially leading to a scenario in which the data term converges independently. Importantly, at this scaling, the data mismatch term does not reduce to a level nearly like that of the wave equation term and declines noticeably slower, indicating that the data mismatch term has fallen into a local minimum. Finally, the convergence processes of the two terms over iterations for four different scaling factors are presented in **Fig. 9**. The results consistently demonstrate that for the model discussed in this paper, iterations fewer than 100 are sufficient. Excessive iterations have resulted in a very slight degree of overfitting.

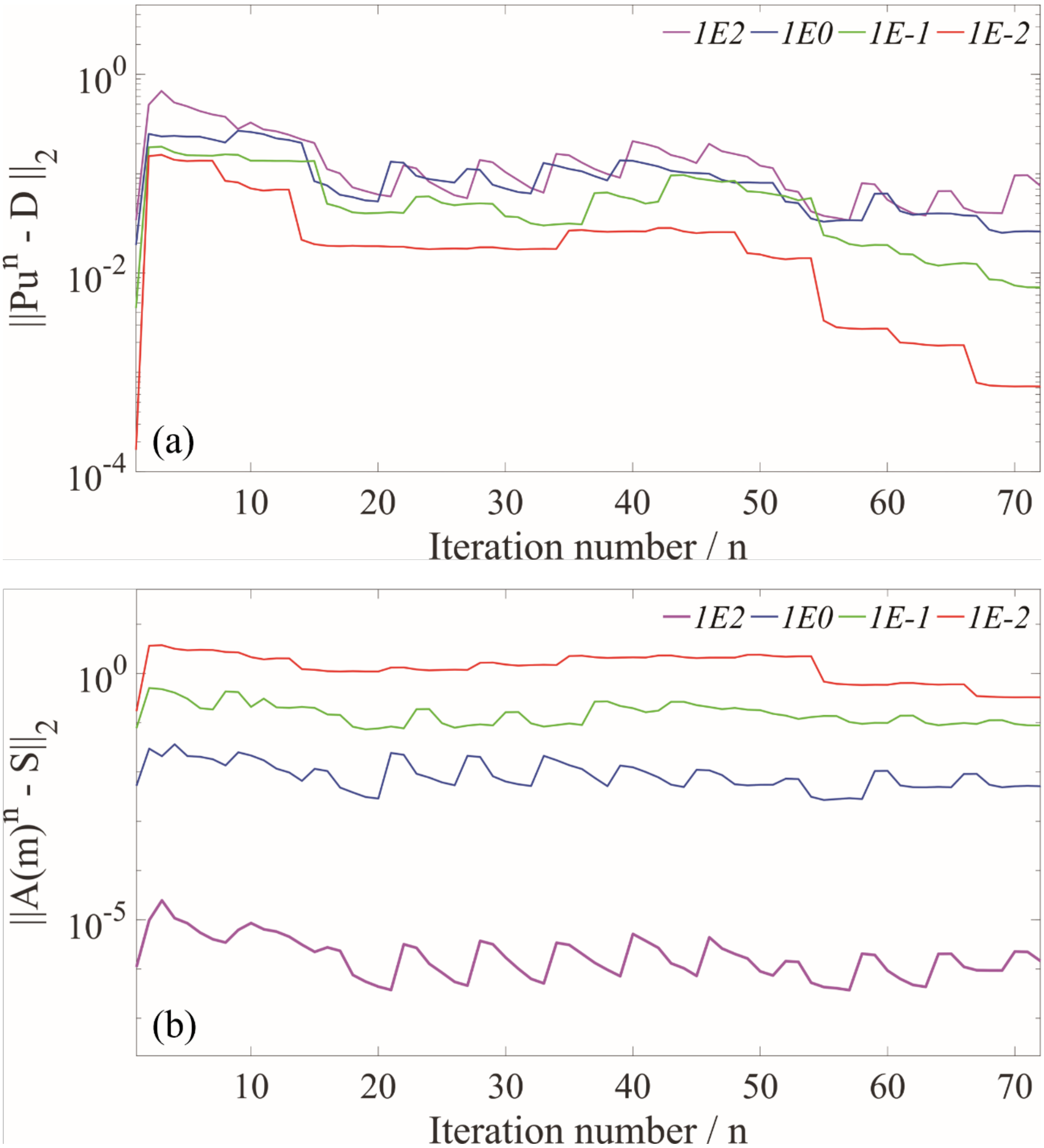

**Fig. 9.** Sensitivity analysis of different parameters $\tau$. (a) Convergence process of the data mismatch penalty term of the objective function. (b) Convergence process of the wave equation penalty term of the objective function. Construction of the objective function according to Eq. **(5)**. The curves are shown for four different scaling levels, $\gamma = 1e-2$ in red, $\gamma = 1e-1$ in green, $\gamma = 1e0$ in blue and $\gamma = 1e2$ in magenta. Additionally, the larger the regularization parameter, the more pronounced the loss spike phenomenon becomes. For the specific model discussed, iterations fewer than 100 are sufficient. Excessive iterations have resulted in a very slight degree of overfitting.

#### *3.1.3. Test I: Inversion and interpretation*

During the inversion phase, we use the limited-memory Broyden–Fletcher–Goldfarb–Shanno (L-BFGS) algorithm (Liu and Nocedal, 1989). Compared to the steepest descent (SD) and conjugate gradient methods (CG), L-BFGS converges faster. While the SD method is simple and updates in the direction of the current gradient, which is often suboptimal and prone to local minima, the CG method, which is based on the conjugacy of two gradients, depends on the condition number and other factors. In contrast, L-BFGS utilizes curvature information (approximating the Hessian matrix), offering a balanced optimization approach in terms of memory usage, convergence speed, and adaptability, particularly suited for large-scale optimization challenges. In brief, L-BFGS can more effectively identify the correct descent direction, thus accelerating convergence.

Subsequently, we present the FWI results obtained using three different algorithms as shown in **Fig. 10**.

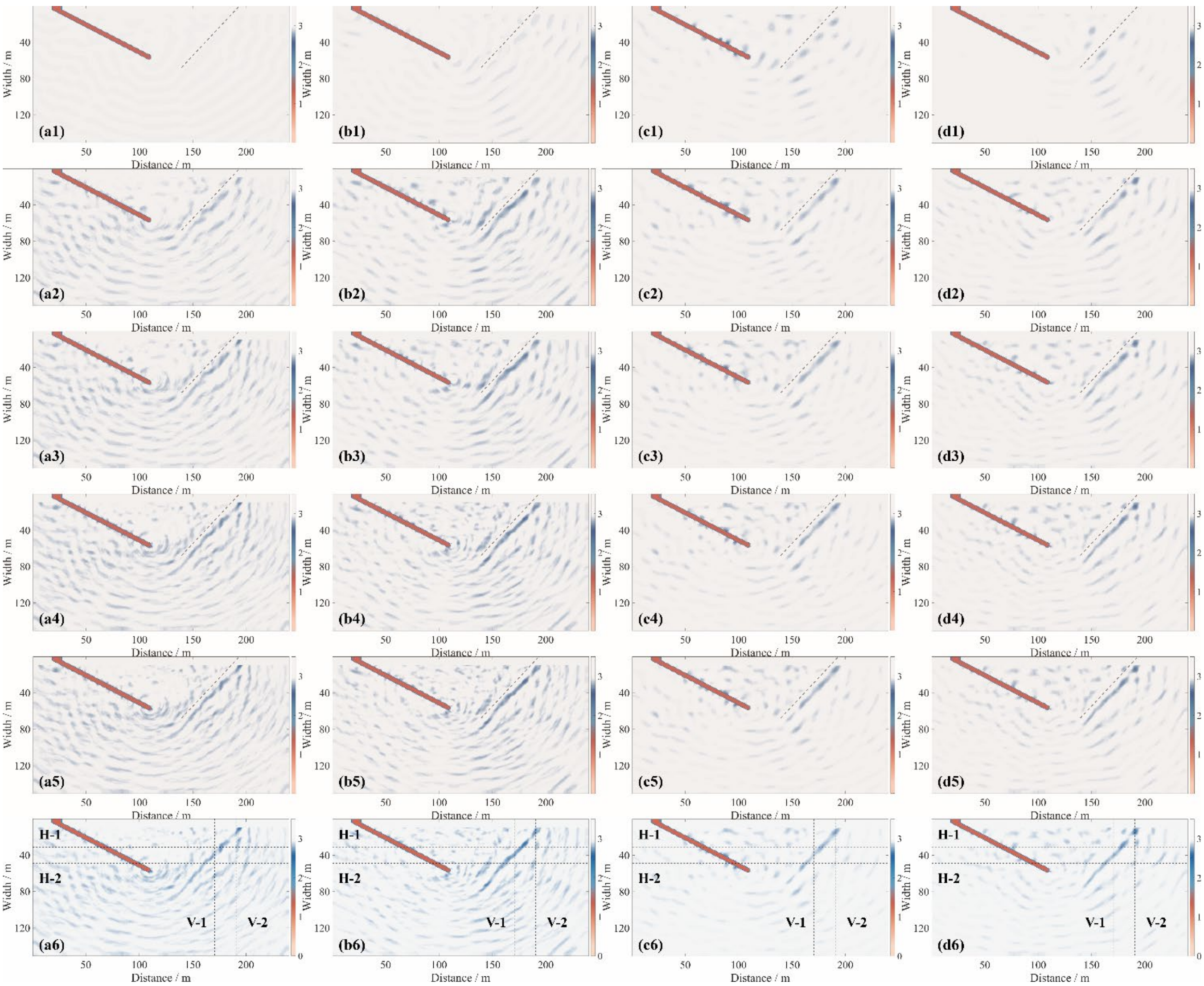

**Fig. 10.** Inversion results comparison. (a1)-(a5) Tikhonov - FWI regularization at 2.0 Hz, 2.7 Hz, 3.2 Hz, 5.2 Hz, and 7.6 Hz. (b1)-(b5) TV - FWI in the same frequency range. (c1)-(c5) Proposed method in the same frequency range and $\gamma = 1e-1$. (d1)-(d5) Proposed method in the same frequency range but $\gamma = 1e0$ case. Horizontal (a6)-(d6) are copies from (a5)-(d5). The black dashed lines indicate the target locations for subsequent velocity profile analyses. V-1 and V-2 represent two sets of vertical sample traces located at 170 m and 190 m, respectively, while H-1 and H-2 represent two sets of horizontal sample traces located at 29 m and 45 m, respectively.

Where **Fig. 10**(a1)-(a5) are the inversion results from Tikhonov regularization-based FWI at frequencies of 2.0 Hz, 2.7 Hz, 3.2 Hz, 5.2 Hz, and 7.6 Hz. **Fig. 10**(b1)-(b5) depict the results from TV regularization-based FWI at the same frequencies, **Fig. 10**(c1)-(c5) display the outcomes from the proposed method-based FWI at $\gamma = 1e-1$, also at the same frequencies, and **Fig. 10**(d1)-(d5) display the outcomes from the proposed method-based FWI at $\gamma = 1e0$, also at the same frequencies. All results utilize the same colour map. From the inversion results of the three different algorithms, it is evident that the Tikhonov regularization-based FWI can identify fault structure, but performs poorly in the inversion of deep, uniform media, exhibiting significant and numerous artefacts, which are theoretically caused by noise, cycle-skipping, and local minima, especially over large areas of homogeneous media. Given that the inversion frequency in FWI should not be set too high and considering the critical importance of low-frequency data for large areas of homogeneous media, the impact of high-wavenumber artefacts must be considered when effective low-frequency data is often unavailable. Furthermore, the choice of regularization and its parameters is crucial; if the regularization parameter is set too low, it complicates the inversion of fault structures and overly smooths the inversion features. Therefore, this also demonstrates that for nonlinear optimization problems, an appropriate choice of regularization is extremely important. Additionally, the inversion results based on TV regularization, as shown in the middle column of **Fig. 10**, clearly exhibit more pronounced and discontinuous artefacts, theoretically due to the using of L1 norm-based TV regularization. Although the sparse constraint of the L1 norm excels in noise reduction while preserving image edges and structural information, crucial for maintaining clear geological interfaces in geophysical inversion, its non-differentiability at the origin can lead to numerical instability in the optimization process, particularly in contexts involving inherently smooth or homogeneous media. Sparse constraints tend to promote gradient sparsity, potentially introducing unnecessary gradient variations within uniform regions, thereby generating artefacts that manifest as unrealistic boundaries or structures. In contrast, the results from the improved algorithm proposed in this paper, as shown in the right two columns of **Fig. 10**, appear more stable, with clearer inversion of faults and tunnel structures. Importantly, they exhibit superior performance in the inversion of uniform media, the inversion results without apparent artefact structures, thus avoiding misinterpretation. Theoretically, this advantage is due to the superiority of the quadratic penalty method and Sobolev space norms. Specifically, the calculations of the wavefield and adjoint operators in the algorithm proposed in this paper have been modified.

Furthermore, the $W_p^1$ regularization utilized in this example, being a non-L1 norm class, ensures differentiability at the origin, and while enhancing regularization effects, avoids excessive artefact generation. Horizontal **Fig. 10**(a6)-(d6) are copies from **Fig. 10**(a5)-(d5). The black dashed lines indicate the target locations for subsequent velocity profile analyses. V-1 and V-2 represent two sets of vertical sample traces located at 170 m and 190 m, respectively, while H-1 and H-2 represent two sets of horizontal sample traces located at 29 m and 45 m, respectively.

Additionally, we also present the model increments of three algorithms at the thirtieth iteration in **Fig. 11**. Since the optimization phase of FWI involves iterative cycling, the model increments at each iteration directly influence the inversion results.

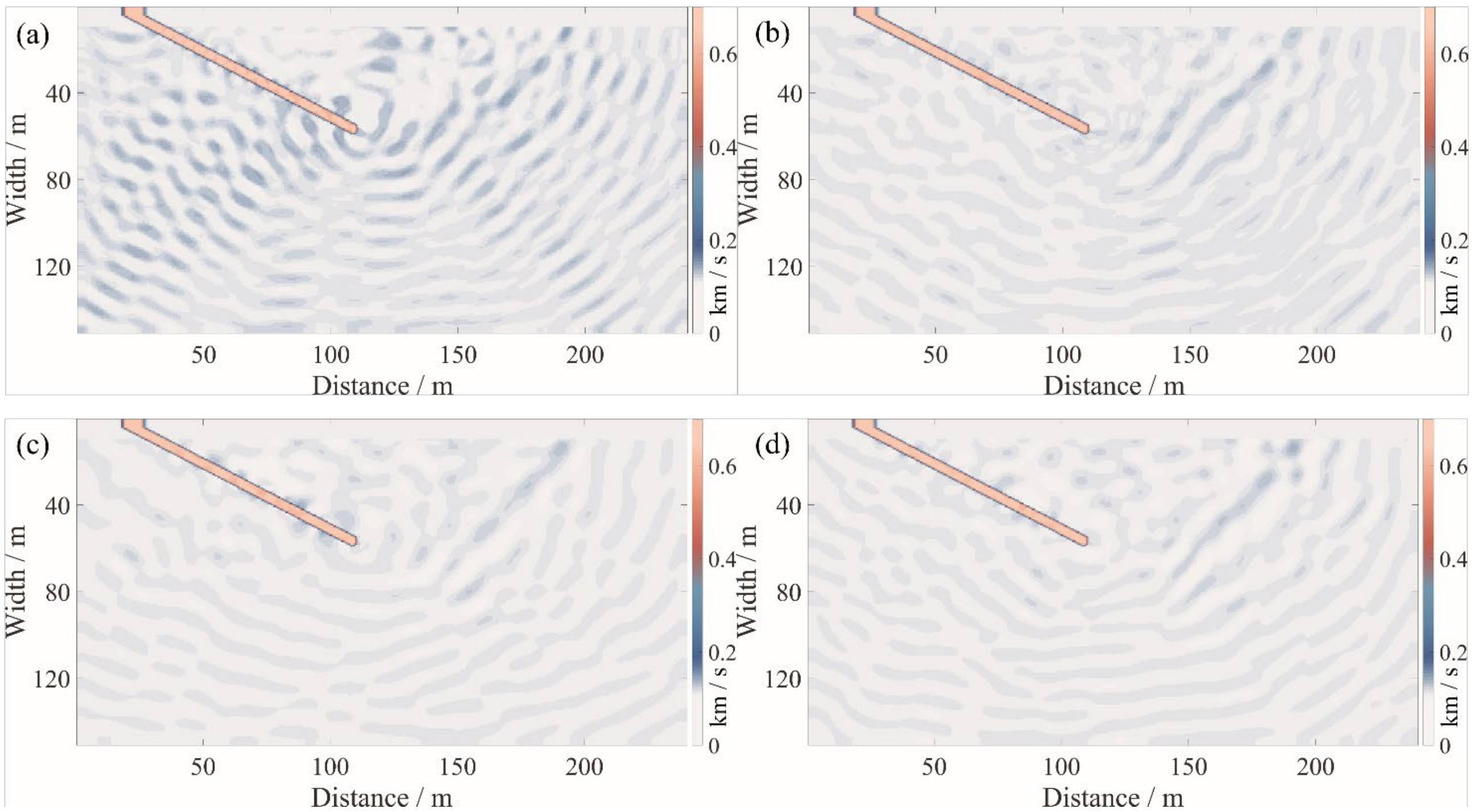

**Fig. 11.** Model increments at the 30th iteration. (a) Tikhonov regularization. (b) TV regularization. (c) Proposed modified FWI and $\gamma = 1e-1$. (d) Proposed modified FWI and $\gamma = 1e0$.

These artefacts significantly reduce the interpretability of the inversion results, especially given their high amplitude, making it challenging to distinguish them from actual geological structures. **Fig. 11**(a) illustrates the model increment for the Tikhonov regularization algorithm. **Fig. 11**(b) displays the model increment based on TV regularization, which shows fewer high-velocity artefacts. **Fig. 11**(c) shows the model increment at the 30th inversion iteration using the improved algorithm proposed at $\gamma = 1\text{e}-1$ in this paper. **Fig. 11**(d) shows the model increment at the 30th inversion iteration using the improved algorithm proposed at $\gamma = 1\text{e}0$. Notably, the improved algorithm results do not exhibit significant artefacts in the deeper parts of the model and provide higher accuracy in the inversion of faults, which is attributed to the soft constrained objective function with a penalty term and a new convex optimization formulation proposed in this paper. Theoretically, in optimization problems like FWI, the original problem can exhibit ill-conditioned behaviour due to its nonlinearity, leading to instability or imprecision in the solution process. For conventional constrained optimization problems, such as standard FWI, directly handling constraints can complicate the solution process, especially when the constraints are complex or nonlinear. Incorporating constraints through a penalty term into the objective function transforms a constrained problem into a soft-constrained one, thereby simplifying the solution process. This approach imposes a high cost on solutions that violate constraints, compelling the solver to find solutions that satisfy the constraints, and increase the condition number, thus improving the numerical stability and convergence of the solution process.

Finally, we present the comparisons of the one-dimensional velocity profiles from three different algorithms. **Fig. 12** compares the lateral velocities at the depth of 29 m (H-1) and 45 m (H-2) within the model, where the grey dashed line represents the initial model velocity curve, the black solid line is the benchmark velocity curve, the red solid line corresponds to the Tikhonov regularization-based velocity curve, the blue solid line is from the TV regularization FWI, the red solid line represents the improved algorithm at $\gamma = 1\text{e}-1$ velocity curve, and the red solid line represents the improved algorithm at $\gamma = 1\text{e}0$ velocity curve.

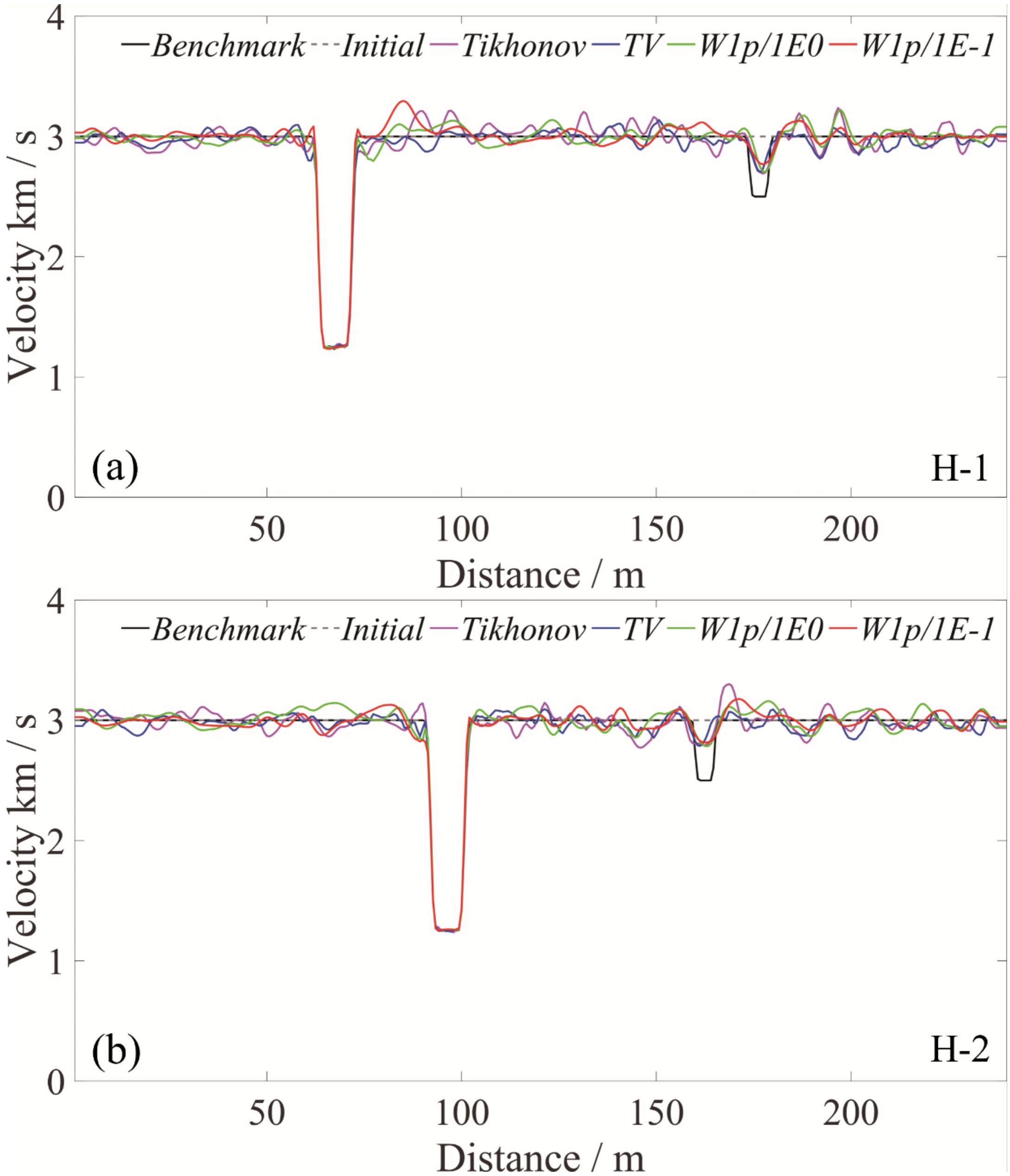

**Fig. 12.** Horizontal velocity profile analysis. (a) 29 m. (b) 45 m. The grey dotted line represents the initial velocity. The black dashed line represents the velocity of the benchmark; the magenta solid line represents the Tikhonov regularization FWI; the blue solid line represents the TV regularization FWI; the red solid line denotes the $\gamma = 1e-1$ proposed modified FWI results in the velocity profile; the green solid line denotes the $\gamma = 1e0$ proposed modified FWI results in the velocity profile. The closer it approaches the black dashed line, the better the inversion results.

In **Fig. 12**(a), a distinct tunnel velocity and another sharp velocity protrusion are observed, with other regions displaying unchanged velocity. It is evident that although the conventional algorithms can identify the location of structural bodies, their performance outside these structures is poor, particularly with sharp spikes and discontinuities at the model's two sides, indicating algorithmic instability. In contrast, the improved algorithm not only accurately identifies fault and tunnel locations but also demonstrates remarkable stability in the uniform velocity regions compared to the other two conventional algorithms. The conclusion of **Fig. 12**(b) is consistent with **Fig. 12**(a). Additionally, **Fig. 13**(a) and **Fig. 13**(b) display the comparison of two sets of vertical velocity slices at 170 m (V-1) and 190 m (V-2) along the model's longitudinal axis.

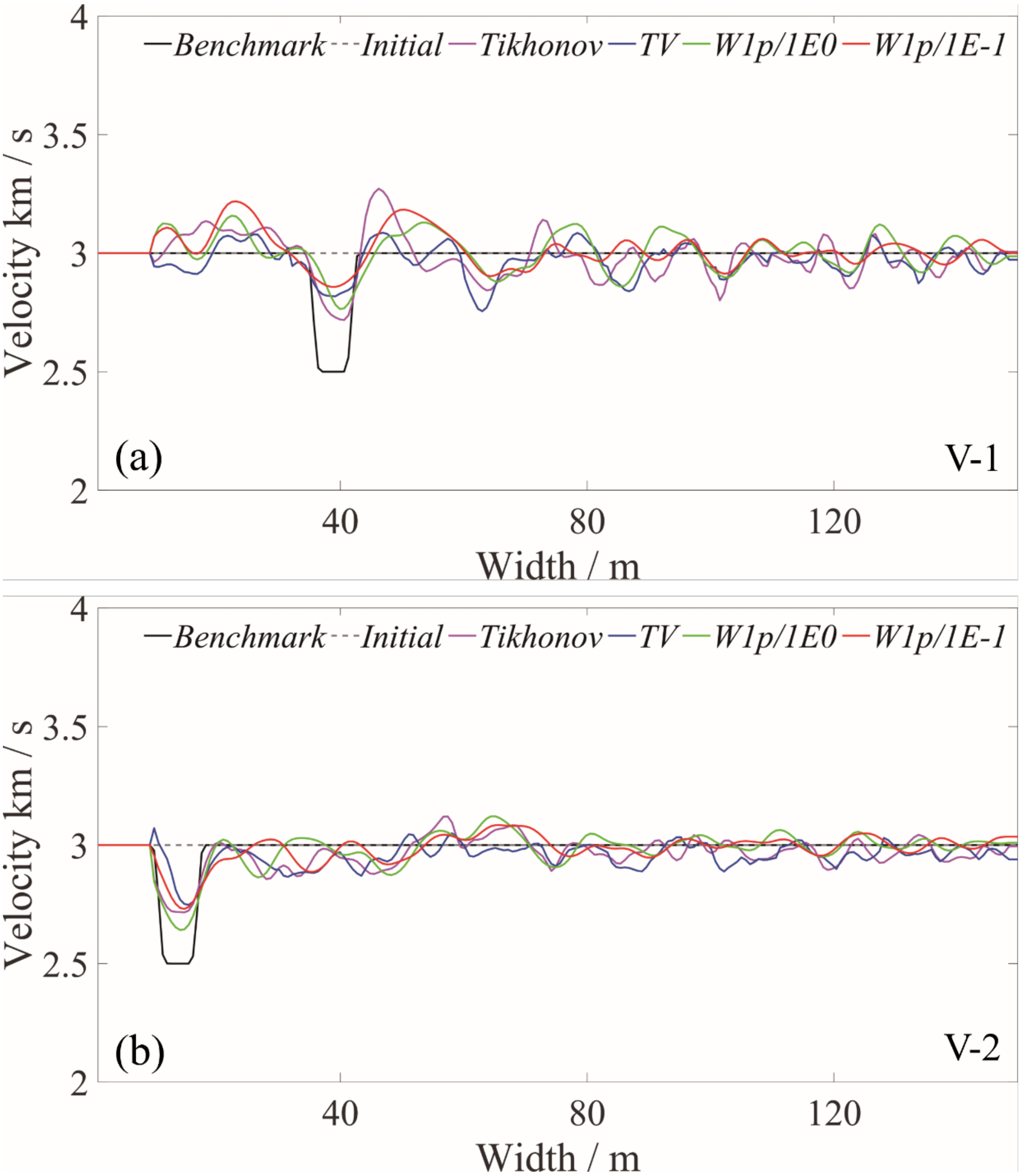

**Fig. 13.** Vertical velocity profile analysis. (a) 170 m. (b) 190 m. The grey dotted line represents the initial velocity. The black dashed line represents the velocity of the benchmark; the magenta solid line represents the Tikhonov regularization FWI; the blue solid line represents the TV regularization FWI; the red solid line denotes the $\gamma = 1\mathrm{e}-1$ proposed modified FWI results in the velocity profile; the green solid line denotes the $\gamma = 1\mathrm{e}0$ proposed modified FWI results in the velocity profile. The closer it approaches the black dashed line, the better the inversion results.

Similar to the conclusions drawn from the previous figure, the proposed algorithm continues to demonstrate superior stability and interpretability, which is evident in its accurate fitting of fault structures and stable inversion of the surrounding rock velocities. Theoretically, by incorporating a quadratic penalty function, we expand the search space, and the optimal value obtained from the penalty function is always less than or equal to the original problem's optimal value, allowing us to interpret the penalty function as a relaxation of the original problem. By introducing a penalty function into FWI, we can transform a constrained problem into a soft-constrained one. The changing of the misfit structure not only simplifies the mathematical formulation and the solving process but also provides greater flexibility to adjust the intensity of the penalty terms, balancing the relationship between the objective function and constraint violations. The flexibility of the new misfit item is particularly important as it allows us to adjust the performance according to the characteristics of the data and the performance of the solving algorithm. In traditional FWI without a penalty function, the model may develop unrealistic physical properties (such as velocity anomalies) under conditions of poor data quality or incomplete data coverage. By incorporating penalty terms, the algorithm is compelled to avoid these non-physical solutions, enhancing the reliability and physical interpretability of the inversion results.

#### *3.1.4. Test II: Numerical simulation for single fault model with free surface and without noise*

In the previous example, we evaluated the inversion performance of the improved algorithm under noisy conditions. For comparison, we also conducted a test in a noise-free setting with a free surface to assess the inversion effectiveness and capability of the algorithm proposed in this study under ideal and noise-free conditions, as shown in **Fig. 14**.

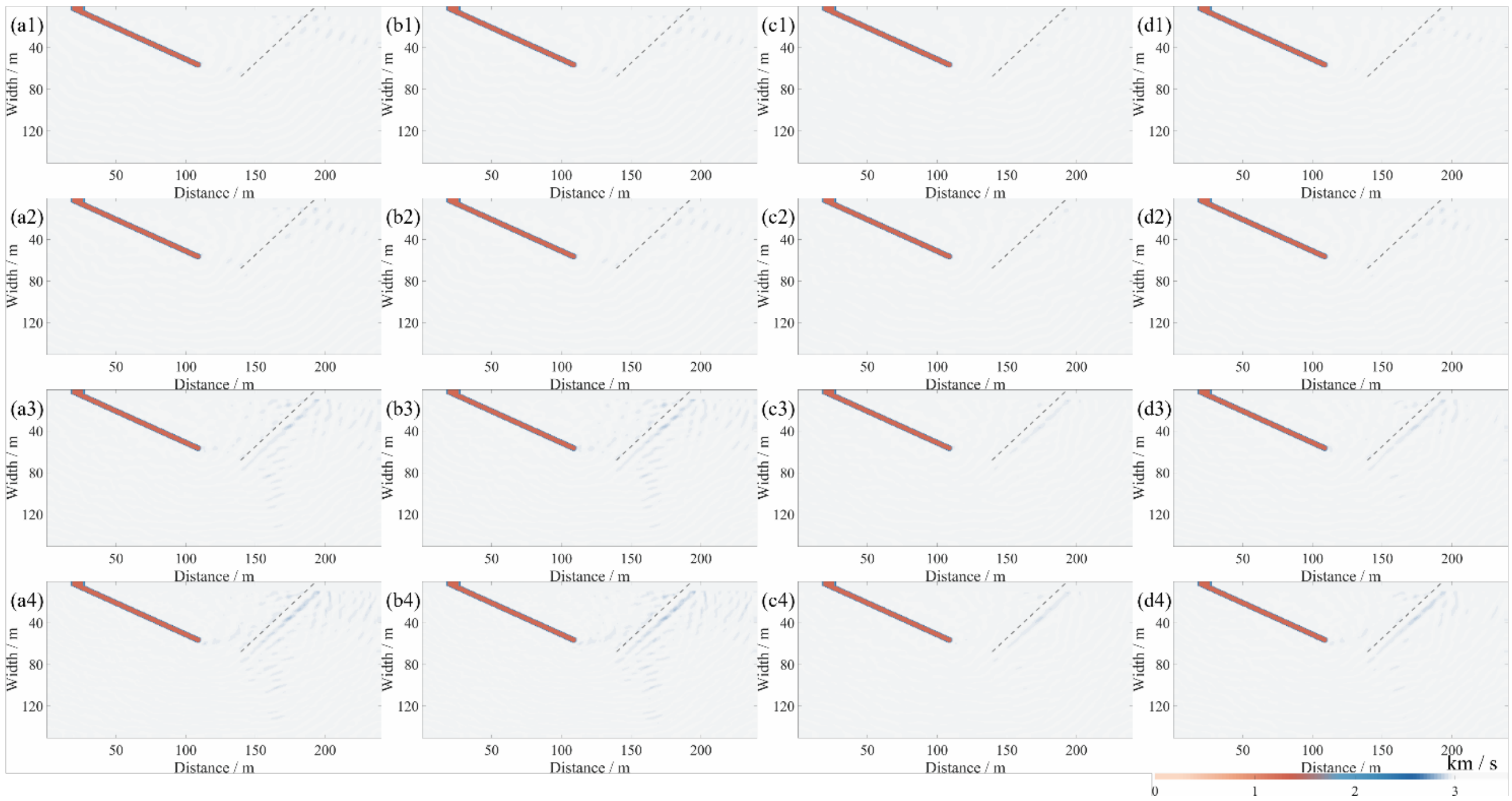

**Fig. 14.** Inversion results in comparison for single fault model without noise. (a1)-(a4) Tikhonov - FWI regularization at 3.0 Hz, 3.3 Hz, 4.4 Hz, and 7.8 Hz. (b1)-(b4) TV - FWI in the same frequency range. (c1)-(c4) Proposed method in the same frequency range and $\gamma = 1\mathrm{e}-1$. (d1)-(d4) Proposed method in the same frequency range but $\gamma = 1\mathrm{e}0$ case.

The results of **Fig. 14** indicate that all methods successfully identify fault structures. However under ideal conditions with a free surface and no noise, the inversion results from the two conventional algorithms still display minor and localized artefacts. The above observation suggests that the inversion results of the two conventional regularization algorithms are nearly identical in the absence of denoising requirements. This is attributed to the fact that the two conventional algorithms differ only slightly in their regularization terms, while other steps, including the computation of the adjoint wavefield, are nearly identical. As a result, the two algorithms produce almost indistinguishable results in scenarios without noise. In contrast, the improved algorithm, especially when $\gamma = 1\mathrm{e}-1$, produces inversion results with almost no artifacts. Such performance demonstrates that the improved algorithm achieves superior inversion outcomes even without the need for denoising. The advantage arises from its distinctive approach to calculating the adjoint wavefield, which deviates from the conventional adjoint wavefield computation formula, effectively confirming the efficiency and superiority of Eq. **(7)**.

## 3.2. Double fault model

To further test the effectiveness, practicality and wide applicability of the algorithm proposed in this paper, we simulate and test the proposed algorithm in other sets of models with double faults. Compared to a single-fault model structure, a double faults scenario presents increased complexity in seismic wave interpretation. The transmission and refraction of seismic waves after the first fault result in very weak reflections at the second fault, with continuous energy attenuation. These weak reflection signals are difficult to identify in seismic records, thus complicating geological structure interpretation. Additionally, changes in wave velocity cause deviations in the wave propagation path. These deviations lead to discrepancies between the expected and actual positions of the second fault, resulting in identification errors. Consequently, recognizing a double-fault structure is significantly more challenging than a single-fault structure, requiring higher algorithm stability and resolution.

### *3.2.1. Test IV: Double Fault Model with free surface and with noise*

First, a model of a double faults structure with a width of 5 m is constructed. As illustrated in **Fig. 15**(a), the two faults are nearly parallel, each with a width of 5 m.

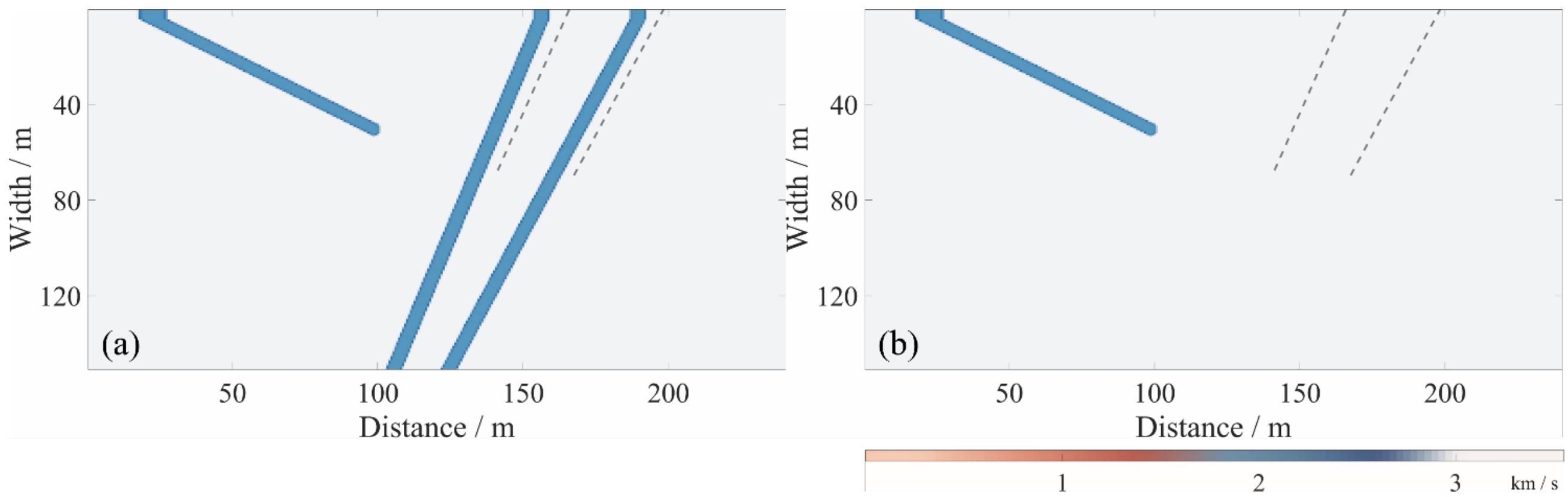

**Fig. 15.** Modelling of a dual fault structure for tunnel advance prediction. (a) Benchmark of the model with two dip-slip faults. (b) The initial model of the benchmark. The grey dotted lines indicate the location of the faults. All two faults have a width of 5 m.

The surrounding rock has a velocity of 3.0 km / s, while the faults have velocities of 2 km / s. The velocity of the tunnel is 2 km / s. The dip angles of the two faults are 65 degrees and 70 degrees, respectively. The top of this model is a free surface, and the overall dimensions of the model are 240 m in width and 140 m in depth. Additionally, a new acquisition system is designed for this model. Ten sources are placed at the tunnel face. Seven equally spaced receivers numbered 1 to 7, are positioned at the tunnel bottom. Furthermore, 25 equally spaced receivers, numbered 8 to 32, are set up at the top of the model. **Fig. 15**(b) shows the initial velocity model. Given that the tunnel structure is known, the complete tunnel structure is included in this initial model. Subsequently, the wavefields for the sources and receivers are presented in **Fig. 16**. **Fig. 16**(a) shows the combined receiver data for receivers numbered 1-7, 8-14 and 15-39. **Fig. 16**(b) illustrates the data with 4.9 *dB* random noise. **Fig. 16**(c) displays the noisy data.

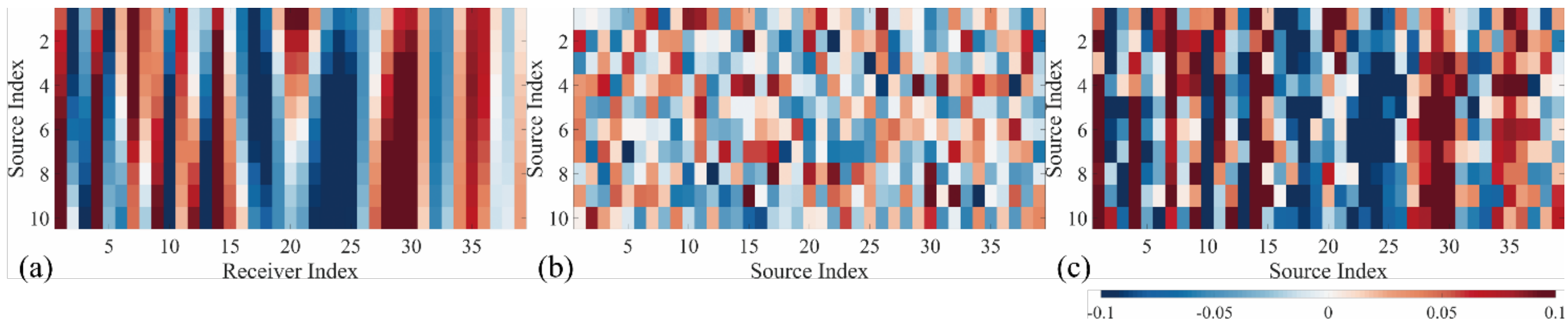

**Fig. 16.** Source receiver data information for the dual faults tunnel model. (a) The real part of the clean data without noise component. The vertical coordinate explicit acquisition system consists of ten seismic sources. The horizontal coordinates from 1 to 7 for receivers on the tunnel floor, the 8 to 14 are the receivers on the tunnel roof and the 8 to 32 receivers are on the free surface. (b) The real part of the random noise. (c) The real part of the data with random noise as a means for further inversion.

Subsequently, we compared the true data corresponding to the 5th source with the simulated data from two algorithms, as illustrated in **Fig. 17**. **Fig. 17**(a) shows the real parts of the data comparisons for receivers numbered 1-7, 8-14 and 15-39, while **Fig. 17**(b) presents the corresponding imaginary parts.

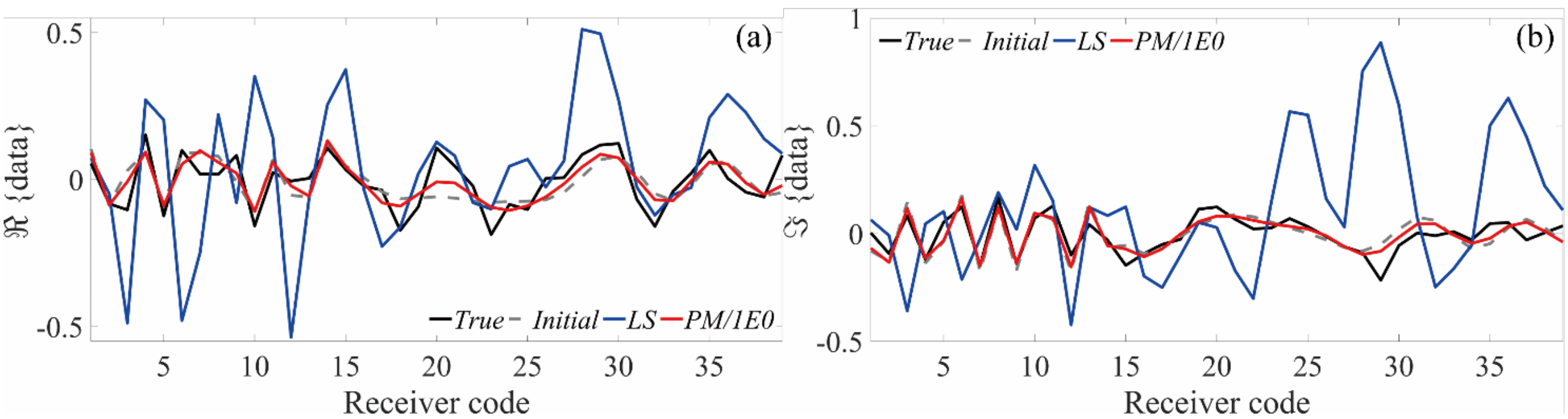

**Fig. 17.** Data for the 5th source and the 1st iteration. (a) The real part of the data received by receivers No.1-39. (b) The imaginary part of the data received by receivers No.1-39. Since the LS simulated data has a significant abnormal amplitude and shift from the observed data, we consider its inversion to be difficult.

In the above figures, the black solid line represents the true data, the grey dashed line corresponds to the data from the initial model, the blue solid line denotes the simulated data after one iteration of the LS method, and the red solid line indicates the simulated data after one iteration of the proposed algorithm. The data after one iteration based on the LS method exhibits a significant abnormal amplitude and shift compared to the observed data, indicating that subsequent inversions would be very challenging. In contrast, the simulated data from one iteration of the proposed algorithm closely matches the observed data and is not constrained by the initial model. The results suggest that the improved algorithm, supported by an expanded search space, facilitates easier inversion and is less dependent on the initial model.

The inversion results of three algorithms are presented in **Fig. 18**. **Fig. 18**(a1)-(a4) show the inversion results based on Tikhonov regularization FWI, **Fig. 18**(b1)-(b4) depict TV regularization FWI's inversion results, and **Fig. 18**(c1)-(c4) illustrate the inversion results based on the proposed improved algorithm.

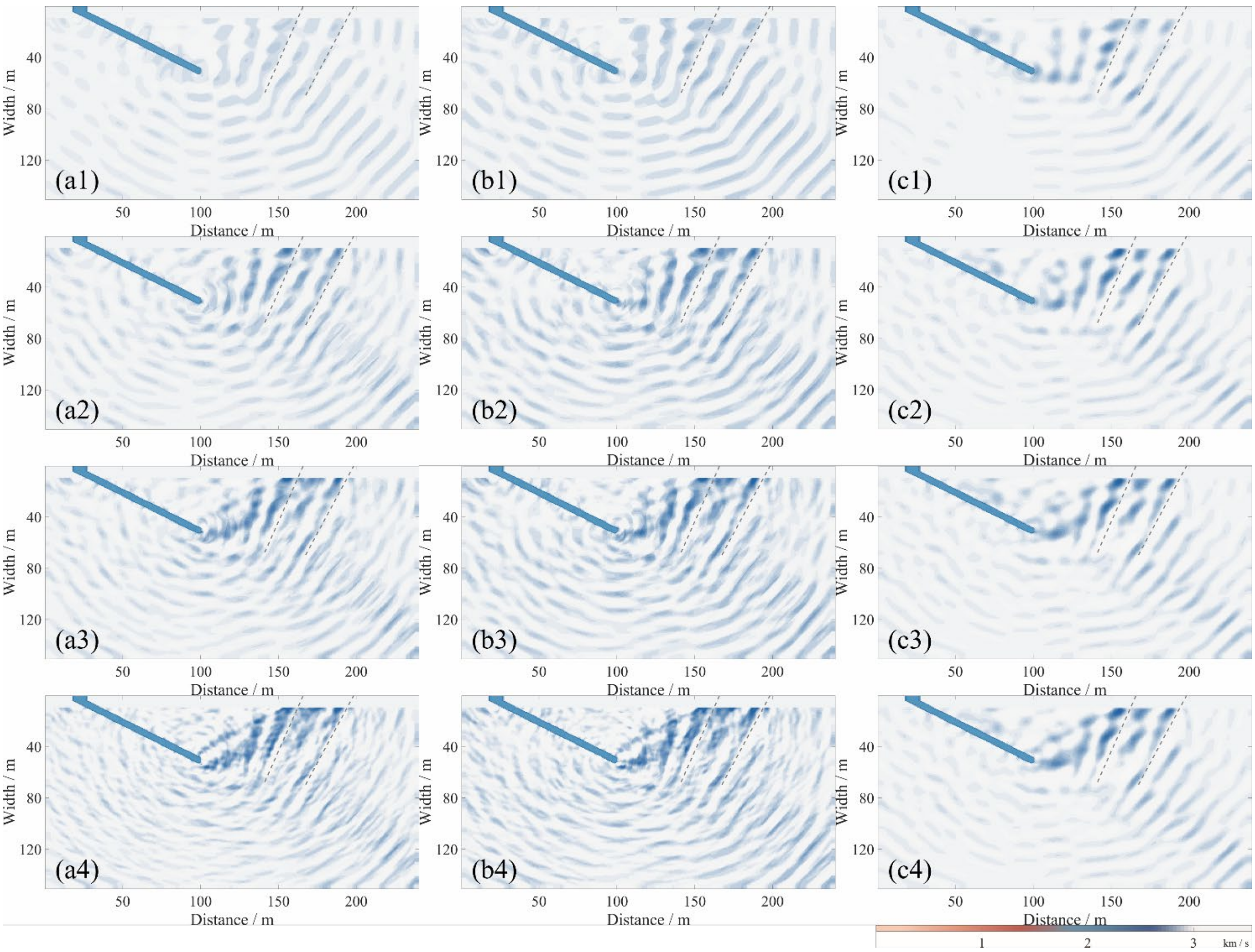

**Fig. 18.** Inversion results comparison. (a1)-(a3) Tikhonov-FWI regularization at 2.0 Hz, 2.7 Hz, 3.5 Hz and 7.5 Hz. (b1)-(b3) TV-FWI in the same frequency range. (c1)-(c3) Proposed method in the same frequency range and $\gamma = 1e0$. The position of the grey auxiliary lines corresponds to that in the true model.

The results of **Fig. 18** indicate that the two conventional algorithms exhibit very blurred fault identification, and due to the significant influence of noise, distinguishing the target structure from artefacts is challenging. In contrast, the improved algorithm demonstrates better denoising capabilities. Compared to conventional algorithms, anomalies are less pronounced, and the simulating effect for the first fault is excellent. However, the inversion effect for the second fault is generally moderate due to reasons such as the shadowing effect of the first fault, which weakens the signal reflected from the second fault or causes it to be obscured by the reflection signal from the first fault. Although the improved algorithm shows certain enhancements over conventional algorithms for the second fault, there remains room for further improvement.

### *3.2.2. Test V: Double Fault Model with free surface and without noise*

Subsequently, we continue to test the inversion results of the three algorithms in the case of a double fault with free surfaces and no noise. The inversion results are shown in **Fig. 19**.

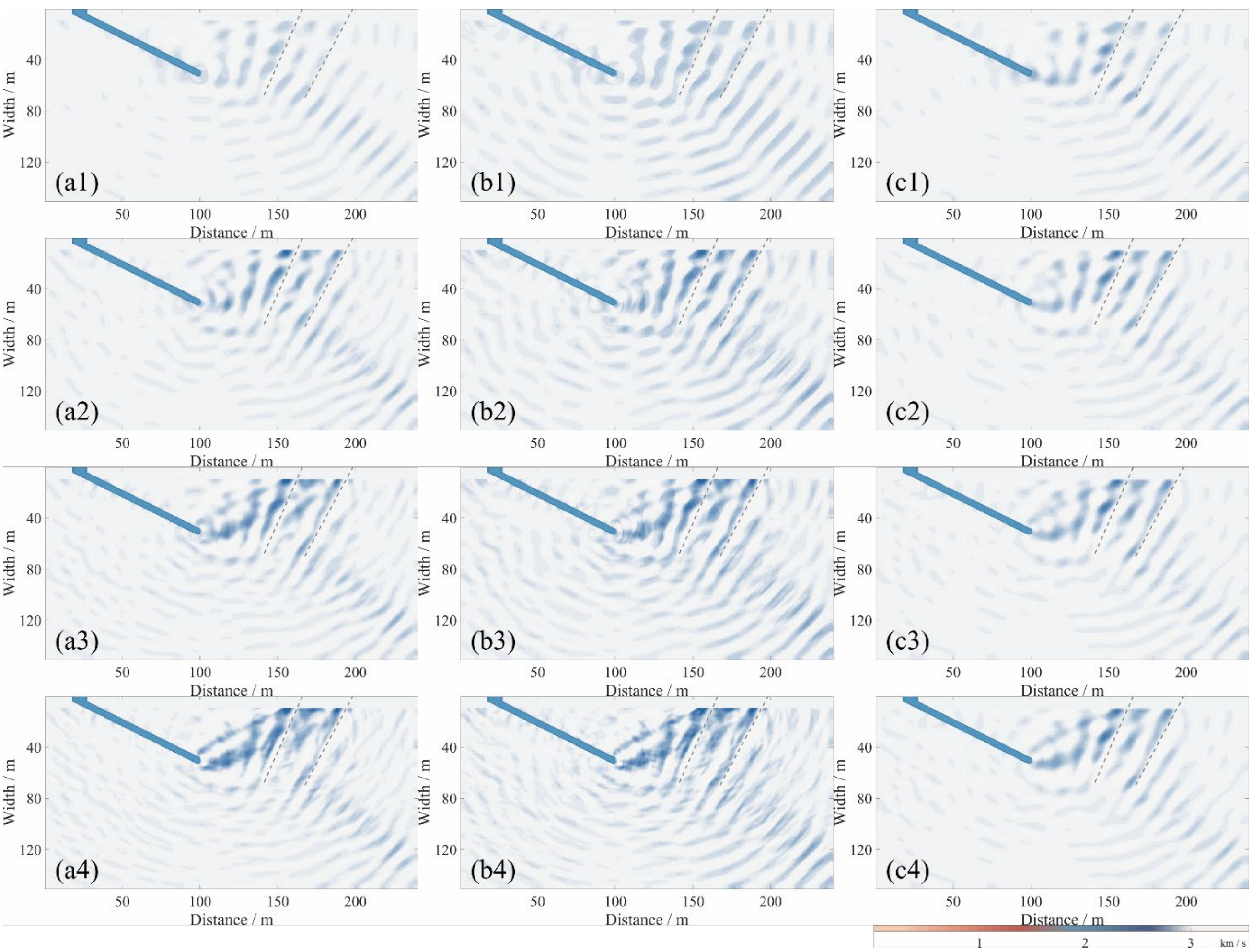

**Fig. 19.** Inversion results comparison. (a1)-(a3) Tikhonov-FWI regularization at 2.0 Hz, 2.7 Hz, 3.5 Hz and 7.5 Hz. (b1)-(b3) TV-FWI in the same frequency range. (c1)-(c3) Proposed method in the same frequency range and $\gamma = 1e0$. The position of the grey auxiliary lines corresponds to that in the true model.

The results indicate that under noise-free conditions, Tikhonov-FWI and TV-FWI achieve better inversion results than before with fewer artifacts and clearer fault structures. Furthermore, the outcomes of the proposed method demonstrate that the inversion results are highly consistent under both noisy and noise-free conditions, which consistency highlights the effectiveness of the proposed algorithm in suppressing noise, even in scenarios involving dual faults. To better compare the differences between **Fig. 19** and **Fig. 18** and better indicate the denoising effect of the proposed algorithm, we calculated the discrepancies between the inversion results under noisy and noise-free conditions. The results are presented in **Fig. 20**.

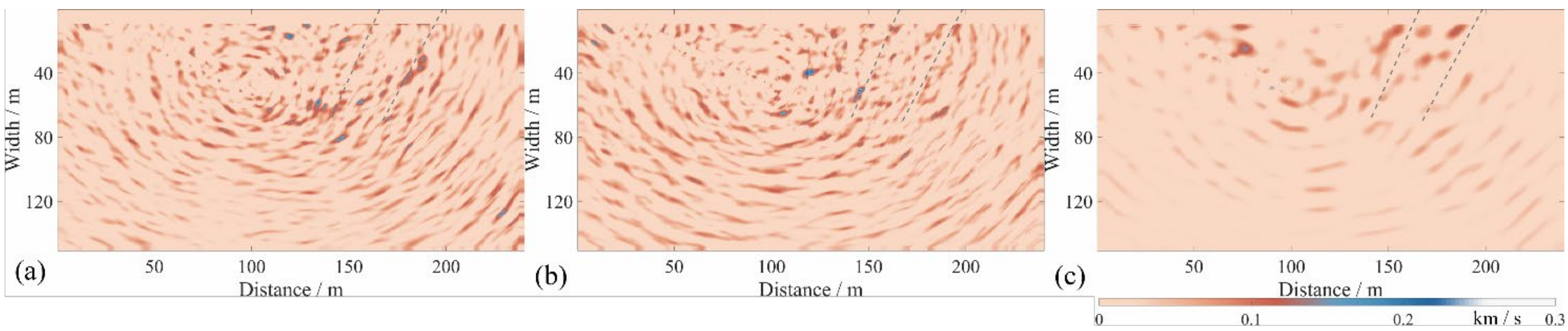

**Fig. 20.** Difference between the inversion results with and without noise. (a) Tikhonov-FWI regularization inversion results's difference. (b) TV-FWI regularization inversion results in difference. (c) Proposed method inversion results's difference.

From **Fig. 20** we can draw a conclusion consistent with the above findings: the proposed algorithm demonstrates superior noise suppression capabilities, which is evidenced by the smaller residuals associated with the improved algorithm in **Fig. 20**, whereas the conventional algorithm exhibits significantly larger residuals. These results provide strong evidence of the enhanced noise suppression performance of the improved algorithm.

## 4. Discussion

This paper introduced the use of an enhanced FWI algorithm for high-resolution inversion and imaging in advance prediction of tunnel constructions. FWI, originally developed for hydrocarbon exploration, has proven to be an effective and widely applicable inversion method in the resource exploration field. The fundamental theory and algorithmic framework of FWI also demonstrate its excellence as an inversion strategy, making it worthy of application in tunnel engineering projects. However, several specific issues need to be addressed and algorithmic improvements made to overcome particular challenges and difficulties, thereby accelerating the promotion and application of FWI in engineering problems.

Compared to large-scale, wide azimuth oil and gas exploration, especially in tunnel prediction problems, the data acquisition system can be significantly restricted, leading to incomplete seismic signal collection. Structures under tunnels, particularly those obscured by air cavities, often block the underlying structures, making weak P-wave reflection signals difficult to capture (Chen et al., 2021), the same problem happens in the experiments of this paper. Therefore, it is necessary to design specific data acquisition systems tailored to particular geological structures to capture signal responses in targeted areas. One of FWI's advantages is that it does not restrict the placement of sources and receivers to tunnel environments; configurations outside the tunnel are also acceptable, significantly enhancing flexibility, especially under small aperture conditions.

Furthermore, in the exploration and inversion of small-scale models for tunnel advance prediction, the role of high wavenumber signals requires greater consideration. At this time, the proposed FWI in this paper does not heavily depend on the initial model, nor does it rely extensively on low-frequency sampling, which is extremely beneficial for FWI. Although high wavenumber signals are sufficient to support FWI for tunnel advance prediction, this does not mean that low frequencies are entirely unnecessary (Bretaudeau et al., 2014). Given that it is a nonlinear optimization problem, caution against local minima is still needed. Moreover, low-frequency waves have better penetration capabilities, allowing for the detection of deeper underground structures; therefore, adopting a multi-scale inversion strategy is necessary (Alani et al., 2020).

Subsequently, considering that the LS method is sensitive to noise and outliers, it is necessary to consider a more robust sparse constraint (Farahmand et al., 2011; Li et al., 2024a). Although an ideal state without any additional noise can achieve an ideal inversion result, in actual production, the noise issue is a critical and indispensable factor affecting the inversion results. This is also why strong noise was introduced in the tests conducted in this article. Conventional FWI algorithms, under large-scale conditions with the support of wide azimuth, multiple source blending, and other enhancements, can effectively improve the SNR (Krebs et al., 2009). Additionally, in scenarios not emphasizing immediacy, conventional pre-processing of pre-stack data, such as filtering techniques and Radon transforms, can be used before proceeding with FWI (Feng, 2020). However, there is not enough time to preprocess during the prediction of structures ahead of tunnel excavation, thus posing higher demands on FWI. Therefore, it is recommended to use enhanced techniques, firstly altering the optimization process and reconstructing the objective function to expand the search space and mitigate the issues caused by noise-induced local minima. Moreover, adopting more reasonable and flexible sparse constraints is also essential; beyond conventional TV regularization, it is advised to use more versatile regularization algorithms, including Huber regularization, K-support regularization, and a series of composite regularization algorithms to flexibly address noise issues (Li et al., 2024a).

Finally, considering that FWI is relatively expensive computationally compared to other conventional methods, the issue of computational cost still needs to be addressed. While transforming second-order optimization into first-order can reduce complexity, it also brings about slower convergence rates. Therefore, it is suggested to use more common optimization acceleration algorithms, such as the alternating direction method of multipliers (Aghamiry et al., 2022).

In summary, after appropriately optimizing and improving FWI, it is very feasible and practical to apply it broadly and inexpensively in tunnel engineering projects.

## 5. Conclusion

We attempt to apply an enhanced FWI based on a penalty method to the advance prediction of tunnel structures. Initially, in response to the limited space available in tunnel construction sites, a combined internal and external acquisition system is utilized, deploying only a few receivers inside the tunnel while placing the majority on the free surface above the tunnel. This combination provides a broader monitoring range and more comprehensive geological information, thereby enhancing the accuracy and timeliness of predictions. Further, research indicates that FWI, as a highly nonlinear optimization problem, exacerbates the high possibility of struggling in local minima with small-scale structures, making it particularly important and challenging to counteract local minima and cycle-skipping issues in tunnel advance prediction scenarios. Consequently, an improved FWI algorithm is proposed, incorporating the quadratic penalty techniques and the multi-scale inversion strategy. By utilizing convex relaxation techniques to transform hard constraints into soft constraints, the efficiency and flexibility of solving optimization problems are enhanced. The misfit function consists of two components: a data misfit term and a term driven by the wave equation, thus minimizing the model and wavefield to expand the search space, alleviating the nonlinearity of inversion and achieving high-resolution inversion. Experimental results also demonstrate that the improved algorithm is not sensitive to the initial model. Subsequently, a Sobolev space regularization algorithm is specifically introduced to address potential random noise issues during engineering exploration. By adjusting different regularization parameters based on noise amplitude, data characteristics, and model features, the regularization path is flexibly adjusted. Lastly, the proposed improved system is applied to some classical fault tunnel advance prediction models. The improved algorithm is also compared and discussed against two other conventional algorithms, with test results validating the effectiveness and superiority of the proposed algorithm. Both theoretical and experimental evidence confirm that applying the proposed FWI to tunnel advance prediction is highly feasible and accurate.

## Declaration of competing interest

The authors declare that they have no known competing financial interests or personal relationships that could have appeared to influence the work reported in this paper.

## Data availability

Data associated with this research are confidential and cannot be released.

## Acknowledgements

## Appendix A. From Least Square to Lagrangian Multipliers to Penalty Method & Parameter Stability

In this appendix, we elaborate in detail on how to calculate the misfit and gradient update based on the penalty method. The objective of the optimization process for FWI is to compute the minimum value of the misfit function (Leeuwen and Herrmann, 2015):

$$\min_{\mathbf{m},\mathbf{u}} f(\mathbf{m}), \tag{A.1}$$

classic algorithms include the explicit imposition of constraints on the objective function:

$$\min_{\mathbf{m},\mathbf{u}} \|\mathbf{Pu}-\mathbf{D}\|_2^2 \quad \text{s.t.} \quad \mathbf{A}(\mathbf{m})\mathbf{u}=\mathbf{S}, \tag{A.2}$$

at this point, the objective function is described as follows:

$$f(\mathbf{m}) = \left\|\mathbf{D}-\mathbf{PA}(\mathbf{m})^{-1}\mathbf{S}\right\|_2^2, \tag{A.3}$$

a well-known weakness of classic algorithms is their propensity to fall into local minima. Consequently, a full-space algorithm has been proposed, related to the saddle-point problem of the Lagrange method:

$$\max_{\mathbf{v}} \min_{\mathbf{m},\mathbf{u}} \mathcal{L}(\mathbf{m},\mathbf{u},\mathbf{v}), \tag{A.4}$$

here, $\mathbf{v}$ is a linear space of the same dimension as the wavefield space:

$$\mathcal{L}(\mathbf{m},\mathbf{u},\mathbf{v}) = \|\mathbf{Pu}-\mathbf{D}\|_2^2 + \langle \mathbf{v},\ \mathbf{A}(\mathbf{m})\mathbf{u}-\mathbf{S}\rangle, \tag{A.5}$$

the method of Lagrangian multipliers optimizes jointly over the model parameters, wavefield, and $\mathbf{v}$. Due to the sparsity of the Hessian matrix in Lagrangian, it is theoretically feasible to advance gradient optimization to the second order:

$$\nabla\mathcal{L}(\mathbf{m},\mathbf{u},\mathbf{v}) = \begin{pmatrix} \mathcal{L}_{\mathbf{m}} \\ \mathcal{L}_{\mathbf{u}} \\ \mathcal{L}_{\mathbf{v}} \end{pmatrix} = \begin{pmatrix} \mathbf{G}(\mathbf{m},\mathbf{u})^T\mathbf{v} \\ \mathbf{A}(\mathbf{m})^T\mathbf{v}+\mathbf{P}^T(\mathbf{Pu}-\mathbf{D}) \\ \mathbf{A}(\mathbf{m})\mathbf{u}-\mathbf{S} \end{pmatrix}, \tag{A.6}$$

and:

$$\nabla^2 \mathcal{L}(\mathbf{m},\mathbf{u},\mathbf{v}) = \begin{pmatrix} \mathbf{R}(\mathbf{m},\mathbf{u},\mathbf{v}) & \mathbf{K}(\mathbf{m},\mathbf{v})^T & \mathbf{G}(\mathbf{m},\mathbf{u})^T \\ \mathbf{K}(\mathbf{m},\mathbf{v}) & \mathbf{P}^T\mathbf{P} & \mathbf{A}(\mathbf{m})^T \\ \mathbf{G}(\mathbf{m},\mathbf{u}) & \mathbf{A}(\mathbf{m}) & 0 \end{pmatrix}, \qquad \textbf{(A.7)}$$

where:

$$\mathbf{G}(\mathbf{m},\mathbf{u}) = \frac{\partial \mathbf{A}(\mathbf{m})\mathbf{u}}{\partial \mathbf{m}}, \ \mathbf{K}(\mathbf{m},\mathbf{v}) = \frac{\partial \mathbf{A}(\mathbf{m})^T \mathbf{v}}{\partial \mathbf{m}}, \ \mathbf{R}(\mathbf{m},\mathbf{u},\mathbf{v}) = \frac{\partial \mathbf{G}(\mathbf{m},\mathbf{u})^T \mathbf{v}}{\partial \mathbf{m}}, \qquad \textbf{(A.8)}$$

when $\mathbf{A}$ is sparse, these Jacobian matrices are usually sparse and can be analytically computed.

However, the major issue with this algorithm is the considerable difficulty in storing v in memory, which demands high requirements. Therefore, a simplified version of half-space optimization has been proposed:

$$\min_{\mathbf{m},\mathbf{u}} F(\mathbf{m}) = \min_{\mathbf{m},\mathbf{u}} \|\mathbf{Pu} - \mathbf{D}\|_2^2 + \tau \|\mathbf{A}(\mathbf{m})\mathbf{u} - \mathbf{S}\|_2^2, \qquad \textbf{(A.9)}$$

where $\mathbf{u}$ is the solution to the quadratic penalty equation:

$$\begin{pmatrix} \mathbf{P} \\ \tau \mathbf{A}(\mathbf{m}) \end{pmatrix} \mathbf{u} = \begin{pmatrix} \mathbf{D} \\ \tau \mathbf{S} \end{pmatrix}, \qquad \textbf{(A.10)}$$

then:

$$\left(\mathbf{P}^H \mathbf{P} + \tau^2 \mathbf{A}(\mathbf{m})^H \mathbf{A}(\mathbf{m})\right)\mathbf{u} = \mathbf{P}^H \mathbf{D} + \tau^2 \mathbf{A}(\mathbf{m})^H \mathbf{S}, \qquad \textbf{(A.11)}$$

compared to the method of Lagrangian multipliers, the gradient optimization process of the simplified version is easier, and the second order does not involve the $\mathbf{v}$ component:

$$\nabla F = \begin{pmatrix} F_{\mathbf{m}} \\ F_{\mathbf{u}} \end{pmatrix} = \begin{pmatrix} \tau \mathbf{G}(\mathbf{m},\mathbf{u})^T (\mathbf{A}(\mathbf{m})\mathbf{u} - \mathbf{S}) \\ \mathbf{P}^T(\mathbf{Pu} - \mathbf{D}) + \tau \mathbf{A}(\mathbf{m})^T (\mathbf{A}(\mathbf{m})\mathbf{u} - \mathbf{S}) \end{pmatrix}, \qquad \textbf{(A.12)}$$

and:

$$\nabla^2 F = \begin{pmatrix} F_{\mathbf{m},\mathbf{m}} & F_{\mathbf{m},\mathbf{u}} \\ F_{\mathbf{u},\mathbf{m}} & F_{\mathbf{u},\mathbf{u}} \end{pmatrix}, \qquad \textbf{(A.13)}$$

where:

$$F_{\mathbf{m},\mathbf{m}} = \tau(\mathbf{G}(\mathbf{m},\mathbf{u})^T\mathbf{G}(\mathbf{m},\mathbf{u}) + \mathbf{R}(\mathbf{m},\mathbf{u},\mathbf{A}(\mathbf{m})\mathbf{u} - \mathbf{S}), \tag{A.14}$$

$$F_{\mathbf{m},\mathbf{u}} = \tau(\mathbf{K}(\mathbf{m},\mathbf{A}(\mathbf{m})\mathbf{u} - \mathbf{S}) + \mathbf{A}(\mathbf{m})^T\mathbf{G}(\mathbf{m},\mathbf{u})), \tag{A.15}$$

$$F_{\mathbf{u},\mathbf{u}} = \mathbf{P}^T\mathbf{P} + \tau\mathbf{A}(\mathbf{m})^T\mathbf{A}(\mathbf{m}), \tag{A.16}$$

we introduce:

$$\mathbf{u}_\tau(\mathbf{m}) = \arg\min_{\mathbf{u}} F(\mathbf{m},\mathbf{u}), \tag{A.17}$$

and define:

$$\Phi(\mathbf{m}) = F(\mathbf{m},\mathbf{u}_\tau(\mathbf{m})), \tag{A.18}$$

from this, we can derive the closed-form solution of the optimization problem:

$$\mathbf{u}_\tau = \left(\tau\mathbf{A}^T(\mathbf{m})\mathbf{A}(\mathbf{m}) + \mathbf{P}^T\mathbf{P}\right)^{-1}\left(\tau\mathbf{A}^T(\mathbf{m})\mathbf{S} + \mathbf{P}\mathbf{D}\right), \tag{A.19}$$

the above equation corresponds to Eq. **(6)**. And we can get:

$$\mathbf{v}_\tau = \tau\left(\mathbf{A}\mathbf{u}_\tau - \mathbf{S}\right), \tag{A.20}$$

Moreover, according to Eq. **(A.7)**, we can obtain the gradient expression of the Lagrangian function with soft constraints:

$$\begin{pmatrix} \mathbf{R}(\mathbf{m},\mathbf{u},\mathbf{v}) & \mathbf{K}(\mathbf{m},\mathbf{v})^T & \mathbf{G}(\mathbf{m},\mathbf{u})^T \\ \mathbf{K}(\mathbf{m},\mathbf{v}) & \mathbf{P}^T\mathbf{P} & \mathbf{A}(\mathbf{m})^T \\ \mathbf{G}(\mathbf{m},\mathbf{u}) & \mathbf{A}(\mathbf{m}) & 0 \end{pmatrix} \begin{pmatrix} \mathbf{m}_\tau^* - \mathbf{m}^* \\ \mathbf{u}_\tau^* - \mathbf{u}^* \\ \mathbf{v}_\tau^* - \mathbf{v}^* \end{pmatrix} = \begin{pmatrix} \mathcal{L}_{\mathbf{m}}(\mathbf{m}_\tau^*,\mathbf{u}_\tau^*,\mathbf{v}_\tau^*) \\ 0 \\ \tau^{-1}\mathbf{v}^* + \mathcal{O}(\tau^{-2}) \end{pmatrix}, \tag{A.21}$$

we find:

$$\begin{aligned}&(\mathbf{R} - \mathbf{K}^T\mathbf{A}^{-1}\mathbf{G} - \mathbf{G}^T\mathbf{A}^{-T}\mathbf{K} + \mathbf{G}^T\mathbf{A}^{-T}\mathbf{P}^T\mathbf{P}\mathbf{A}^{-1}\mathbf{G})(\mathbf{m}_\tau^* - \mathbf{m}^*) = \\ &\mathcal{L}_{\mathbf{m}}(\mathbf{m}_\tau^*,\mathbf{u}_\tau^*,\mathbf{v}_\tau^*) + \tau^{-1}\mathbf{Q}\mathbf{v}^* + \mathcal{O}(\tau^{-2}),\end{aligned} \tag{A.22}$$

where:

$$\mathbf{Q} = \mathbf{G}^T\mathbf{A}^{-T}\mathbf{P}^T\mathbf{P}\mathbf{A}^{-1} - \mathbf{K}^T\mathbf{A}^{-1}, \tag{A.23}$$

then, after ignoring the second-order term of the reduced Hessian matrix, we get:

$$\mathbf{H} \approx \mathbf{G}^T \mathbf{A}^{-T} \mathbf{P}^T \mathbf{P} \mathbf{A}^{-1} \mathbf{G}, \tag{A.24}$$

and:

$$\mathbf{Q} \approx \mathbf{G}^T \mathbf{A}^{-T} \mathbf{P}^T \mathbf{P} \mathbf{A}^{-1}, \tag{A.25}$$

then, we can find

$$\left\| \mathbf{m}_\tau^* - \mathbf{m}^* \right\|_2 \le \left\| \mathbf{H}^{-1} \right\|_2 (\varepsilon + \tilde{\tau}^{-1} \left\| \mathbf{G}^T \mathbf{A}^{-T} \mathbf{P}^T \right\|_2 \left\| \mathbf{D} - \mathbf{PA}(\mathbf{m}_\tau^*)^{-1} \mathbf{S} \right\|_2), \tag{A.26}$$

where:

$$\mu = \left\| \mathbf{D} - \mathbf{PA}(\mathbf{m}_\tau^*)^{-1} \mathbf{S} \right\|_2, \tag{A.27}$$

represents the data mismatch. The subsequent priority is to introduce the scaling operation

$$\left\| \mathbf{m}_\tau^* - \mathbf{m}^* \right\|_2 \le \left\| \mathbf{H}^{-1} \right\|_2 (\tilde{\varepsilon} + \tilde{\tau}^{-1} \tilde{\mu}), \tag{A.28}$$

based on this, we reconstruct partial model parameters using the principal eigenvectors, rather than a complete reconstruction. We discuss the different scaling levels in detail in the experimental section.

## Figure Captions

**Fig. 1.** Visualization and simulation of the $W_p^1$ regularization. (a) *p* values ranging from 0.5 to 2, the azimuth angle (AZ) is 50, and the elevation angle (EL) is 15, (AZ: 50°, EL: 15°). (b) *p* values ranging from 0.5 to 2, the AZ is negative 50°, and the elevation angle (EL) is 15°, (AZ: -50°, EL: 15°). (c*)* p values ranging from 0.5 to 0.7, which exhibits the characteristics of the 0.5 norm. (d*)* p values ranging from 1 to 1.2, which exhibits the characteristics of TV regularization. (e*)* p values ranging from 1.5 to 1.7, which exhibits the characteristics of the 1.5 norm. (f*)* p value is equal to 2, which exhibits the characteristics of the Tikhonov regularization.

**Fig. 2.** Schematic of the model structure. (a) Tunnel structural model with an inclined fault. (b) The modelling results after tilting the schematic **Fig. 2**(a), the top of the modelling is a free surface to simulate the real ground surface (the number of receivers in the schematic does not represent the real number of receivers).

**Fig. 3.** Modelling of a fault structure for tunnel seismic advance prediction. (a) Benchmark of the model with a tilted fault. (b) The initial model of the benchmark. The tunnel structure was removed for inversion self-consistency testing. The grey dotted line is the indicator line.

**Fig. 4.** Forward wavefield information. (a) The real part of the forward wavefields. (b) The imaginary part of the forward wavefields. (c) The real part of the initial background wavefield. (d) The imaginary part of the background wavefield.

**Fig. 5.** Source-receiver data for the tunnel model. (a) The real part of the clean data without noise component. The vertical coordinate explicit acquisition system consists of two seismic sources, horizontal coordinates from 15 to 39 for receivers on the free surface, and from 1 to 14 for receivers on the tunnel floor and roof. (b) The imaginary part of the clean data without noise component. (c) The real part of the random noise. (d) The imaginary part of the random noise. (e) The real part of the data with random noise as a means for further steps. (f) The imaginary part of the data with random noise as a means for further steps.

**Fig. 6.** Comparison of inversion tests with different scaling factors $\gamma$ for maximum eigenvalues $\mathbf{A}^{-T}\mathbf{P}^{T}\mathbf{P}\mathbf{A}^{-1}$. (a) $\gamma = 1\mathrm{e}-2$. (b) $\gamma = 1\mathrm{e}-1$. (c) $\gamma = 1\mathrm{e}0$. (d) $\gamma = 1\mathrm{e}2$. Directly below the inversion results, a set of vertical velocity contrasts is displayed, where the black solid line corresponds to the benchmark velocity, the red solid line corresponds to the inverted velocity, and the grey dashed line corresponds to the initial velocity. As the scaling factor increases, leading to a smaller reciprocal of the scaling factor, the inversion results tend to become unstable due to the dominance of the data mismatch term. Conversely, with smaller scaling factors, resulting in a larger reciprocal of the scaling factor, the inversion results are more stable as the wave equation term becomes more predominant. However, if the scaling factor is too small, the data mismatch term will be invalidated.

**Fig. 7.** Maximum eigenvalue curve of $\mathbf{A}^{-T}\mathbf{P}^{T}\mathbf{P}\mathbf{A}^{-1}$ in relation to the penalty parameter $\tau$ in Eq. **(8)**. The curves are shown for three different scaling levels, $\gamma = 1\mathrm{e}-2$ in red, $\gamma = 1\mathrm{e}-1$ in green, $\gamma = 1\mathrm{e}0$ in blue and $\gamma = 1\mathrm{e}2$ in magenta.

**Fig. 8.** Relative importance of the data mismatch term and the wave equation penalty term in the initial stage of the inversion. The curves are shown for four different scaling levels, $\gamma = 1\mathrm{e}-2$ in red, $\gamma = 1\mathrm{e}-1$ in green, $\gamma = 1\mathrm{e}0$ in blue and $\gamma = 1\mathrm{e}2$ in magenta. Smaller scaling factors can ensure that both terms maintain convergent behaviour simultaneously. At $\gamma = 1\mathrm{e}-2$, the wavefield term's correlation with the data mismatch term is small, potentially leading to a scenario in which the wave equation item converges independently, suggesting the possibility of the wave equation term falling into a local minimum. At $\gamma = 1\mathrm{e}-1$, the wave equation term converges slightly faster than the data term, but these two items are balanced. At $\gamma = 1\mathrm{e}0$, the data mismatch term converges faster than the wave equation term, but the data misfit item is dominant, wave equation item is auxiliary. However, at $\gamma = 1\mathrm{e}2$, the wavefield term's correlation with the data mismatch term is also too small, potentially leading to a scenario in which the data term converges independently. Importantly, at this scaling, the data mismatch term does not reduce to a level nearly like that of the wave equation term and declines noticeably slower, indicating that the data mismatch term has fallen into a local minimum.

**Fig. 9.** Sensitivity analysis of different parameters $\tau$. (a) Convergence process of the data mismatch penalty term of the objective function. (b) Convergence process of the wave equation penalty term of the objective function. Construction of the objective function according to Eq. **(5)**. The curves are shown for four different scaling levels, $\gamma = 1\mathrm{e}-2$ in red, $\gamma = 1\mathrm{e}-1$ in green, $\gamma = 1\mathrm{e}0$ in blue and $\gamma = 1\mathrm{e}2$ in magenta. Additionally, the larger the regularization parameter, the more pronounced the loss spike phenomenon becomes. For the specific model discussed, iterations fewer than 100 are sufficient. Excessive iterations have resulted in a very slight degree of overfitting.

**Fig. 10.** Inversion results comparison. (a1)-(a5) Tikhonov - FWI regularization at 2.0 Hz, 2.7 Hz, 3.2 Hz, 5.2 Hz, and 7.6 Hz. (b1)-(b5) TV - FWI in the same frequency range. (c1)-(c5) Proposed method in the same frequency range and $\gamma = 1\text{e}-1$ . (d1)-(d5) Proposed method in the same frequency range but $\gamma = 1\text{e}0$ case. Horizontal (a6)-(d6) are copies from (a5)-(d5). The black dashed lines indicate the target locations for subsequent velocity profile analyses. V-1 and V-2 represent two sets of vertical sample traces located at 170 m and 190 m, respectively, while H-1 and H-2 represent two sets of horizontal sample traces located at 29 m and 45 m, respectively.

**Fig. 11.** Model increments at the 30th iteration. (a) Tikhonov regularization. (b) TV regularization. (c) Proposed modified FWI and $\gamma = 1\text{e}-1$ . (d) Proposed modified FWI and $\gamma = 1\text{e}0$ .

**Fig. 12.** Horizontal velocity profile analysis. (a) 29 m. (b) 45 m. The grey dotted line represents the initial velocity. The black dashed line represents the velocity of the benchmark; the magenta solid line represents the Tikhonov regularization FWI; the blue solid line represents the TV regularization FWI; the red solid line denotes the $\gamma = 1\text{e}-1$ proposed modified FWI results in the velocity profile; the green solid line denotes the $\gamma = 1\text{e}0$ proposed modified FWI results in the velocity profile. The closer it approaches the black dashed line, the better the inversion results.

**Fig. 13.** Vertical velocity profile analysis. (a) 170 m. (b) 190 m. The grey dotted line represents the initial velocity. The black dashed line represents the velocity of the benchmark; the magenta solid line represents the Tikhonov regularization FWI; the blue solid line represents the TV regularization FWI; the red solid line denotes the $\gamma = 1\text{e}-1$ proposed modified FWI results in the velocity profile; the green solid line denotes the $\gamma = 1\text{e}0$ proposed modified FWI results in the velocity profile. The closer it approaches the black dashed line, the better the inversion results.

**Fig. 14.** Inversion results in comparison for single fault model without noise. (a1)-(a4) Tikhonov - FWI regularization at 3.0 Hz, 3.3 Hz, 4.4 Hz, and 7.8 Hz. (b1)-(b4) TV - FWI in the same frequency range. (c1)-(c4) Proposed method in the same frequency range and $\gamma = 1\text{e}-1$ . (d1)-(d4) Proposed method in the same frequency range but $\gamma = 1\text{e}0$ case.

**Fig. 15.** Modelling of a dual fault structure for tunnel advance prediction. (a) Benchmark of the model with two dip-slip faults. (b) The initial model of the benchmark. The grey dotted lines indicate the location of the faults. All two faults have a width of 5 m.

**Fig. 16.** Source receiver data information for the dual faults tunnel model. (a) The real part of the clean data without noise component. The vertical coordinate explicit acquisition system consists of ten seismic sources. The horizontal coordinates from 1 to 7 for receivers on the tunnel floor, the 8 to 14 are the receivers on the tunnel roof and the 8 to 32 receivers are on the free surface. (b) The real part of the random noise. (c) The real part of the data with random noise as a means for further

inversion.

**Fig. 17.** Data for the 5th source and the 1st iteration. (a) The real part of the data received by receivers No.1-39. (b) The imaginary part of the data received by receivers No.1-39. Since the LS simulated data has a significant abnormal amplitude and shift from the observed data, we consider its inversion to be difficult.

**Fig. 18.** Inversion results comparison. (a1)-(a3) Tikhonov-FWI regularization at 2.0 Hz, 2.7 Hz, 3.5 Hz and 7.5 Hz. (b1)-(b3) TV-FWI in the same frequency range. (c1)-(c3) Proposed method in the same frequency range and $\gamma = 1e0$. The position of the grey auxiliary lines corresponds to that in the true model.

**Fig. 19.** Inversion results comparison. (a1)-(a3) Tikhonov-FWI regularization at 2.0 Hz, 2.7 Hz, 3.5 Hz and 7.5 Hz. (b1)-(b3) TV-FWI in the same frequency range. (c1)-(c3) Proposed method in the same frequency range and $\gamma = 1e0$. The position of the grey auxiliary lines corresponds to that in the true model.

**Fig. 20.** Difference between the inversion results with and without noise. (a) Tikhonov-FWI regularization inversion results's difference. (b) TV-FWI regularization inversion results in difference. (c) Proposed method inversion results's difference.